\documentclass{nature}

\usepackage{caption}
\usepackage{graphicx}
\usepackage[scaled]{helvet}
\usepackage{amsmath}
\usepackage{bm}
\newcommand*{\myfont}{\fontfamily{phv}\selectfont}
\usepackage{hyperref}

\usepackage{lineno}
\usepackage{color, colortbl}
 \definecolor{LightGrey}{rgb}{0.95,.95,.95}

\title{Abstract representations of events arise from mental errors in learning and memory}

\author{Christopher W. Lynn$^1$, Ari E. Kahn$^{2,3}$, Nathaniel Nyema$^3$, \& Danielle S. Bassett$^{1,3,4,5,6,7,*}$}

\begin{document}


\maketitle

\begin{affiliations}
\item Department of Physics \& Astronomy, College of Arts \& Sciences, University of Pennsylvania, Philadelphia, PA 19104, USA
\item Department of Neuroscience, Perelman School of Medicine, University of Pennsylvania, Philadelphia, PA 19104, USA
\item Department of Bioengineering, School of Engineering \& Applied Science, University of Pennsylvania, Philadelphia, PA 19104, USA
\item Department of Electrical \& Systems Engineering, School of Engineering \& Applied Science, University of Pennsylvania, Philadelphia, PA 19104, USA
\item Department of Neurology, Perelman School of Medicine, University of Pennsylvania, Philadelphia, PA 19104, USA
\item Department of Psychiatry, Perelman School of Medicine, University of Pennsylvania, Philadelphia, PA 19104, USA
\item Santa Fe Institute, Santa Fe, NM 87501, USA
\end{affiliations}

\newpage

\begin{abstract}

Humans are adept at uncovering abstract associations in the world around them, yet the underlying mechanisms remain poorly understood. Intuitively, learning the higher-order structure of statistical relationships should involve complex mental processes. Here we propose an alternative perspective: that higher-order associations instead arise from natural errors in learning and memory. Using the free energy principle, which bridges information theory and Bayesian inference, we derive a maximum entropy model of people's internal representations of the transitions between stimuli. Importantly, our model (i) affords a concise analytic form, (ii) qualitatively explains the effects of transition network structure on human expectations, and (iii) quantitatively predicts human reaction times in probabilistic sequential motor tasks. Together, these results suggest that mental errors influence our abstract representations of the world in significant and predictable ways, with direct implications for the study and design of optimally learnable information sources.

\end{abstract}

\newpage

\section*{Introduction}

Our experience of the world is punctuated in time by discrete events, all connected by an architecture of hidden forces and causes. In order to form expectations about the future, one of the brain's primary functions is to infer the statistical structure underlying past experiences.\cite{Hyman-01, Sternberg-01, Johnson-01} In fact, even within the first year of life, infants reliably detect the frequency with which one phoneme follows another in spoken language.\cite{Saffran-01} By the time we reach adulthood, uncovering statistical relationships between items and events enables us to perform abstract reasoning,\cite{Bousfield-01} identify visual patterns,\cite{Fiser-01} produce language,\cite{Friederici-01} develop social intuition,\cite{Tompson-01} and segment continuous streams of data into self-similar parcels.\cite{Reynolds-01} Notably, each of these functions requires the brain to identify statistical regularities across a range of scales. It has long been known that people are sensitive to differences in individual transition probabilities such as those between words or concepts.\cite{Saffran-01, Fiser-01} Additionally, mounting evidence suggests that humans can also infer abstract (or higher-order) statistical structures, including hierarchical patterns within sequences of stimuli,\cite{Meyniel-01} temporal regularities on both global and local scales,\cite{Dehaene-01} abstract concepts within webs of semantic relationships,\cite{Piantadosi-01} and general features of sparse data.\cite{Tenenbaum-01}

To study this wide range of statistical structures in a unified framework, scientists have increasingly employed the language of network science,\cite{Newman-01} wherein stimuli or states are conceptualized as nodes in a graph with edges or connections representing possible transitions between them. In this way, a sequence of stimuli often reflects a random walk along an underlying transition network,\cite{Gomez-01, Newport-01, Garvert-01} and we can begin to ask which network features give rise to variations in human learning and behavior. This perspective has been particularly useful, for example, in the study of artificial grammars,\cite{Cleeremans-01} wherein human subjects are tasked with inferring the grammar rules (i.e., the network of transitions between letters and words) underlying a fabricated language.\cite{Gomez-02} Complementary research in statistical learning has demonstrated that modules (i.e., communities of densely-connected nodes) within transition networks are reflected in brain imaging data\cite{Schapiro-01} and give rise to stark shifts in human reaction times.\cite{Karuza-01} Together, these efforts have culminated in a general realization that people's internal representations of a transition structure are strongly influenced by its higher-order organization.\cite{Kahn-01, Karuza-03} But how does the brain learn these abstract network features? Does the inference of higher-order relationships require sophisticated hierarchical learning algorithms? Or instead, do natural errors in cognition yield a ``blurry" representation, making the coarse-grained architecture readily apparent?

To answer these questions, here we propose a single driving hypothesis: that when building models of the world, the brain is finely-tuned to maximize accuracy while simultaneously minimizing computational complexity. Generally, this assumption stems from a rich history exploring the trade-off between brain function and computational cost,\cite{Tversky-01, De-01} from sparse coding principles at the neuronal level\cite{Vinje-01} to the competition between information integration and segregation at the whole-brain level\cite{Tononi-01} to the notion of exploration versus exploitation\cite{Cohen-01} and the speed-accuracy trade-off\cite{Wickelgren-01} at the behavioral level. To formalize our hypothesis, we employ the free energy principle,\cite{Jaynes-01} which has become increasingly utilized to investigate constraints on cognitive functioning\cite{Ortega-01} and explain how biological systems maintain efficient representations of the world around them.\cite{Friston-01} Despite this thorough treatment of the accuracy-complexity trade-off in neuroscience and psychology, the prevailing intuition in statistical learning maintains that the brain is either optimized to perform Bayesian inference,\cite{Piantadosi-01, Tenenbaum-01} which is inherently error free, or hierarchical learning,\cite{Meyniel-01, Dehaene-01, Newport-01, Cleeremans-01} which typically entails increased rather than decreased computational complexity.

Here, we show that the competition between accuracy and computational complexity leads to a maximum entropy (or minimum complexity) model of people's internal representations of events.\cite{Shannon-01, Jaynes-01}  As we decrease the complexity of our model, allowing mental errors to take effect, higher-order features of the transition network organically come into focus while the fine-scale structure fades away, thus providing a concise mechanism explaining how people infer abstract statistical relationships. To a broad audience, our model provides an accessible mapping from transition networks to human behaviors, with particular relevance for the study and design of optimally learnable transition structures -- either between words in spoken and written language,\cite{Shannon-01, Cleeremans-01, Gomez-02} notes in music,\cite{Brown-01} or even concepts in classroom lectures.\cite{Lake-01}

\section*{Results}

\subsection{Network effects on human expectations.}

\begin{figure}[t!]
\centering
\includegraphics[width = .85\textwidth]{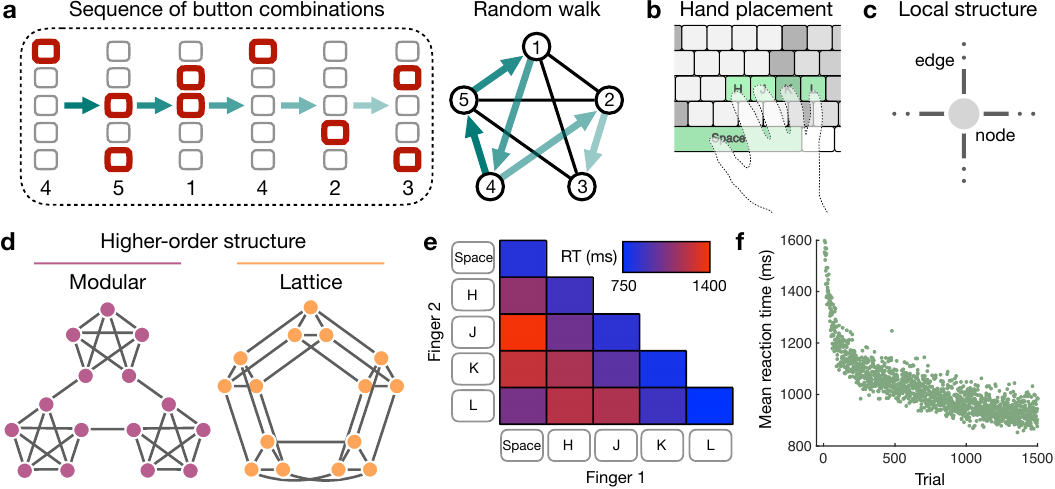} \\
\raggedright
\myfont\textbf{Fig. 1: Subjects respond to sequences of stimuli drawn as a random walk on an underlying transition graph.}
\captionsetup{labelformat=empty}
{\spacing{1.25} \caption{\label{experiment} \myfont \textbf{a}, Example sequence of visual stimuli (left) representing a random walk on an underlying transition network (right). \textbf{b}, For each stimulus, subjects are asked to respond by pressing a combination of one or two buttons on a keyboard. \textbf{c}, Each of the 15 possible button combinations corresponds to a node in the transition network. We only consider networks with nodes of uniform degree $k=4$ and edges with uniform transition probability $0.25$. \textbf{d}, Subjects were asked to respond to sequences of 1500 such nodes drawn from two different transition architectures: a modular graph (left) and a lattice graph (right). \textbf{e}, Average reaction times for the different button combinations, where the diagonal elements represent single-button presses and the off-diagonal elements represent two-button presses. \textbf{f}, Average reaction times as a function of trial number, characterized by a steep drop-off in the first 500 trials followed by a gradual decline in the remaining 1000 trials. In \textbf{e} and \textbf{f}, averages are taken over responses during random walks on the modular and lattice graphs. Source data are provided as a Source Data file.}}
\end{figure}

\noindent In the cognitive sciences, mounting evidence suggests that human expectations depend critically on the higher-order features of transition networks.\cite{Gomez-01, Newport-01} Here, we make this notion concrete with empirical evidence for higher-order network effects in a probabilistic sequential response task.\cite{Kahn-01} Specifically, we presented human subjects with sequences of stimuli on a computer screen, each stimulus depicting a row of five grey squares with one or two of the squares highlighted in red (Fig. \ref{experiment}a). In response to each stimulus, subjects were asked to press one or two computer keys mirroring the highlighted squares (Fig. \ref{experiment}b). Each of the 15 different stimuli represented a node in an underlying transition network, upon which a random walk stipulated the sequential order of stimuli (Fig. \ref{experiment}a). By measuring the speed with which a subject responded to each stimulus, we were able to infer their expectations about the transition structure: a fast reaction reflected a strongly-anticipated transition, while a slow reaction reflected a weakly-anticipated (or surprising) transition.\cite{Hyman-01, Sternberg-01, McCarthy-01, Kahn-01}

While it has long been known that humans can detect differences in transition probabilities -- for instance, rare transitions lead to sharp increases in reaction times\cite{Saffran-01, Fiser-01} -- more recently it has become clear that people's expectations also reflect the higher-order architecture of transition networks.\cite{Schapiro-01, Karuza-01, Karuza-02, Kahn-01} To clearly study these higher-order effects without the confounding influence of variations in transition probabilities, here we only consider transition graphs with a uniform transition probability of $0.25$ on each edge, thereby requiring nodes to have uniform degree $k=4$ (Fig. \ref{experiment}c). Specifically, we consider two different graph topologies: a \textit{modular} graph with three communities of five densely-connected nodes and a \textit{lattice} graph representing a $3\times 5$ grid with periodic boundary conditions (Fig. \ref{experiment}d). Since all transitions across both graphs have uniform probability, any systematic variations in behavior between different parts of a graph, or between the two graphs themselves, must stem from differences in the graphs' higher-order modular or lattice structures.

\begin{figure}[t!]
\centering
\includegraphics[width = .55\textwidth]{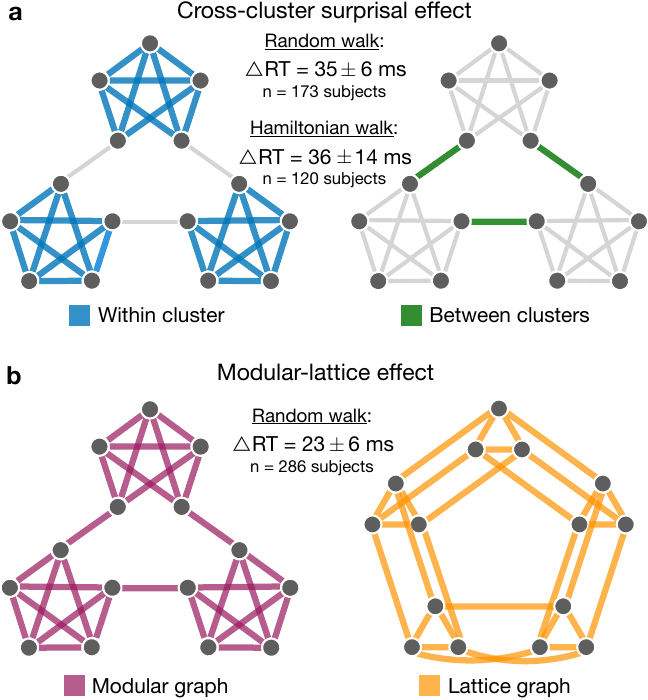} \\
\raggedright
\myfont\textbf{Fig. 2: The effects of higher-order network structure on human reaction times.}
\captionsetup{labelformat=empty}
{\spacing{1.25} \caption{\label{effects} \myfont \textbf{a}, Cross-cluster surprisal effect in the modular graph, defined by an average increase in reaction times for between-cluster transitions (right) relative to within-cluster transitions (left). We detect significant differences in reaction times for random walks ($p < 0.001$, $t = 5.77$, $\text{df} = 1.61 \times 10^5$) and Hamiltonian walks ($p = 0.010$, $t = 2.59$, $\text{df} = 1.31\times 10^4$). For the mixed effects models used to estimate these effects, see Supplementary Tables 1 and 3. \textbf{b}, Modular-lattice effect, characterized by an overall increase in reaction times in the lattice graph (right) relative to the modular graph (left). We detect a significant difference in reaction times for random walks ($p < 0.001$, $t = 3.95$, $\text{df} = 3.33\times 10^5$); see Supplementary Table 2 for the mixed effects model. Measurements were on independent subjects, statistical significance was computed using two-sided $F$-tests, and confidence intervals represent standard deviations. Source data are provided as a Source Data file.}}
\end{figure}

Regressing out the dependence of reaction times on the different button combinations (Fig. \ref{experiment}e), the natural quickening of reactions with time\cite{Baayen-01} (Fig. \ref{experiment}f), and the impact of stimulus recency (see Methods), we identify two effects of higher-order network structure on subjects' reactions. First, in the modular graph we find that reactions corresponding to within-cluster transitions are 35 ms faster than reactions to between-cluster transitions ($p < 0.001$, $F$-test; Supplementary Table 1), an effect known as the \textit{cross-cluster surprisal}\cite{Karuza-02, Kahn-01} (Fig. \ref{effects}a). Similarly, we find that people are more likely to respond correctly for within-cluster transitions than between-cluster transitions (Supplementary Table 8). Second, across all transitions within each network, we find that reactions in the modular graph are 23 ms faster than those in the lattice graph ($p < 0.001$, $F$-test; Supplementary Table 2), a phenomenon that we coin the \textit{modular-lattice effect} (Fig. \ref{effects}b).

Thus far, we have assumed that variations in human behavior stem from people's internal expectations about the network structure. However, it is important to consider the possible confound of stimulus recency: the tendency for people to respond more quickly to stimuli that have appeared more recently.\cite{Murdock-01, Baddeley-01} To ensure that the observed network effects are not simply driven by recency, we performed a separate experiment that controlled for recency in the modular graph by presenting subjects with sequences of stimuli drawn according to Hamiltonian walks, which visit each node exactly once.\cite{Schapiro-01} Within the Hamiltonian walks, we still detect a significant cross-cluster surprisal effect (Fig. \ref{effects}a; Supplementary Tables 3-5). Additionally, we controlled for recency in our initial random walk experiments by focusing on stimuli that previously appeared a specific number of trials in the past. Within these recency-controlled data, we find that both the cross-cluster surprisal and modular-lattice effects remain significant (Supplementary Figs. 2 and 3). Finally, for all of our analyses throughout the paper we regress out the dependence of reaction times on stimulus recency (see Methods). Together, these results demonstrate that higher-order network effects on human behavior cannot be explained by recency alone.

In combination, our experimental observations indicate that people are sensitive to the higher-order architecture of transition networks. But how do people infer abstract features like community structure from sequences of stimuli? In what follows, we turn to the free energy principle to show that a possible answer lies in understanding the subtle role of mental errors.

\subsection{Network effects reveal errors in graph learning.}

As humans observe a sequence of stimuli or events, they construct an internal representation $\hat{A}$ of the transition structure, where $\hat{A}_{ij}$ represents the expected probability of transitioning from node $i$ to node $j$. Given a running tally $n_{ij}$ of the number of times each transition has occurred, one might na\"{i}vely expect that the human brain is optimized to learn the true transition structure as accurately as possible.\cite{Stachenfeld-01, Momennejad-01} This common hypothesis is represented by the maximum likelihood estimate,\cite{Boas-01} taking the simple form
\begin{equation}
\label{MLE}
\hat{A}^{\text{MLE}}_{ij} = \frac{n_{ij}}{\sum_k n_{ik}}.
\end{equation}
To see that human behavior does not reflect maximum likelihood estimation, we note that Eq. (\ref{MLE}) provides an unbiased estimate of the transition structure;\cite{Boas-01} that is, the estimated transition probabilities in $\hat{A}^{\text{MLE}}$ are evenly distributed about their true value $0.25$, independent of the higher-order transition structure. Thus, the fact that people's reaction times depend systematically on abstract features of the network marks a clear deviation from maximum likelihood estimation. To understand how higher-order network structure impacts people's internal representations, we must delve deeper into the learning process itself.

Consider a sequence of nodes $(x_1, x_2,\hdots)$, where $x_t\in \{1,\hdots,N\}$ represents the node observed at time $t$ and $N$ is the size of the network (here $N = 15$ for all graphs). To update the maximum likelihood estimate of the transition structure at time $t+1$, one increments the counts $n_{ij}$ using the following recursive rule,
\begin{equation}
\label{C1}
n_{ij}(t+1) = n_{ij}(t) + \left[i = x_t\right]\left[j = x_{t+1}\right],
\end{equation}
where the Iverson bracket $\left[\cdot\right] = 1$ if its argument is true and 0 otherwise. Importantly, we note that at each time $t+1$, a person must recall the previous node that occurred at time $t$; in other words, they must associate a cause $x_t$ to each effect $x_{t+1}$ that they observe. While maximum likelihood estimation requires perfect recollection of the previous node at each step, human errors in perception and recall are inevitable.\cite{Gregory-01, Howard-01, Howard-03} A more plausible scenario is that, when attempting to recall the node at time $t$, a person instead remembers the node at time $t - \Delta t$ with some decreasing probability $P(\Delta t)$, where $\Delta t \ge 0$. This memory distribution, in turn, generates an internal belief about which node occurred at time $t$,
\begin{equation}
\label{B}
B_t(i) = \sum_{\Delta t = 0}^{t-1} P(\Delta t) \left[i = x_{t - \Delta t}\right].
\end{equation}
Updating Eq. (\ref{C1}) accordingly, we arrive at a learning rule that accounts for natural errors in perception and recall,
\begin{equation}
\label{C2}
\tilde{n}_{ij}(t+1) = \tilde{n}_{ij}(t) + B_t(i)\left[j = x_{t+1}\right].
\end{equation}
Using this revised counting rule, we can begin to form more realistic predictions about people's internal estimates of the transition structure, $\hat{A}_{ij} = \tilde{n}_{ij}/\sum_k \tilde{n}_{ik}$.

We remark that $P(\Delta t)$ does not represent the forgetting of past stimuli altogether; instead, it reflects the local shuffling of stimuli in time. If one were to forget past stimuli at some fixed rate -- a process that is important for some cognitive functions\cite{Richards-01} -- this would merely introduce white noise into the maximum likelihood estimate $\hat{A}^{\text{MLE}}$ (see Supplementary Discussion). By contrast, we will see that, by shuffling the order of stimuli in time, people are able to gather information about the higher-order structure of the underlying transitions.

\subsection{Choosing a memory distribution: The free energy principle.}

\noindent In order to make predictions about people's expectations, we must choose a particular mathematical form for the memory distribution $P(\Delta t)$. To do so, we begin with a single driving hypothesis: that the brain is finely-tuned to (i) minimize errors and (ii) minimize computational complexity. Formally, we define the error of a recalled stimulus to be its distance in time from the desired stimulus (i.e., $\Delta t$), such that the average error of a candidate distribution $Q(\Delta t)$ is given by $E(Q) = \sum_{\Delta t} Q(\Delta t)\Delta t$. By contrast, it might seem difficult to formalize the computational complexity associated with a distribution $Q$. Intuitively, we would like the complexity of $Q$ to increase with increasing certainty. Moreover, as a first approximation we expect the complexity to be approximately additive such that the cost of storing two independent memories equals the costs of the two memories themselves. As famously shown by Shannon, these two criteria of monotonicity and additivity are sufficient to derive a quantitative definition of complexity\cite{Shannon-01} -- namely, the negative entropy $-S(Q) = \sum_{\Delta t} Q(\Delta t)\log Q(\Delta t)$.

Together, the total cost of a distribution $Q$ is its free energy $F(Q) = \beta E(Q) - S(Q)$, where $\beta$ is the inverse temperature parameter, which quantifies the relative value that the brain places on accuracy versus efficiency.\cite{Ortega-01} In this way, our assumption about resource constraints in the brain leads to a particular form for $P$: it should be the distribution that minimizes $F(Q)$, namely the Boltzmann distribution\cite{Jaynes-01}
\begin{equation}
\label{Boltzmann}
P(\Delta t) = \frac{1}{Z}e^{-\beta\Delta t},
\end{equation}
where $Z$ is the normalizing constant (see Methods). Free energy arguments similar to the one presented here have been used increasingly to formalize constraints on cognitive functions,\cite{Friston-01, Ortega-01} with applications from active inference\cite{Friston-04} and Bayesian learning under uncertainty\cite{Friston-01} to human action and perception with temporal or computational limitations.\cite{Ortega-01, Ortega-02, Gershman-02} Taken together, Eqs. (\ref{B}-\ref{Boltzmann}) define our maximum entropy model of people's internal transition estimates $\hat{A}$.

\begin{figure}
\centering
\includegraphics[width = .85\textwidth]{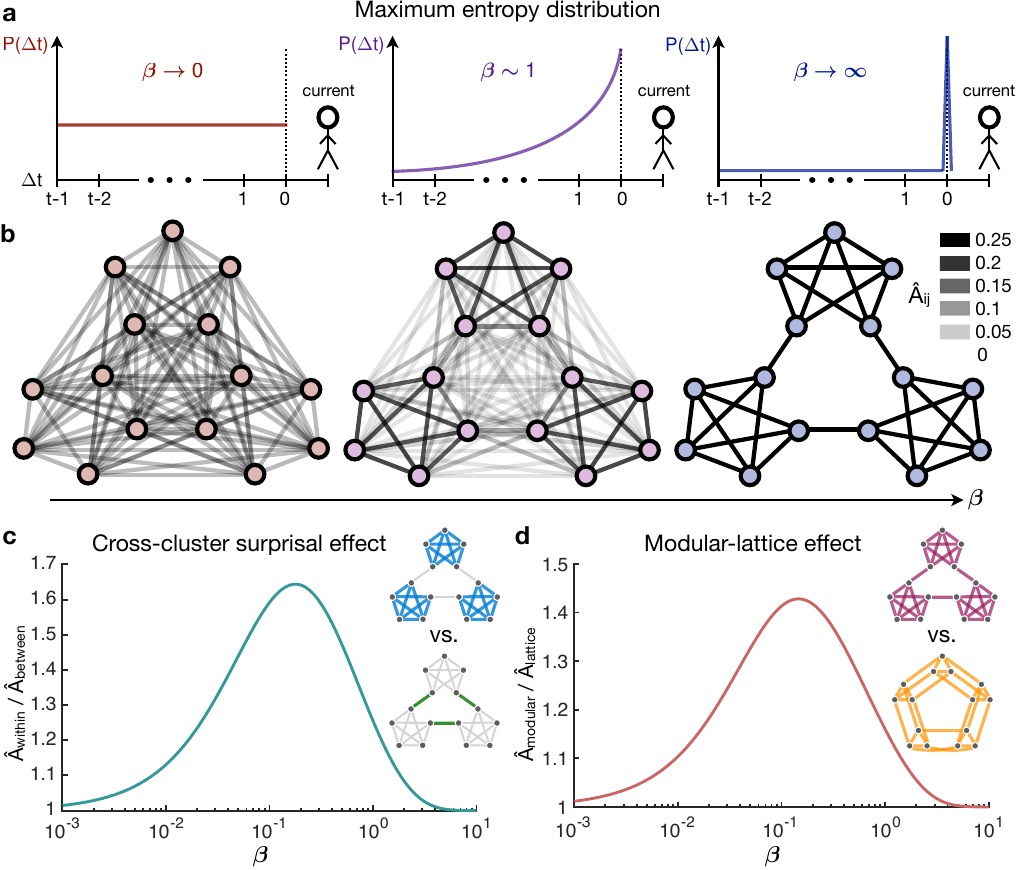} \\
\raggedright
\myfont\textbf{Fig. 3: A maximum entropy model of transition probability estimates in humans.}
\captionsetup{labelformat=empty}
{\spacing{1.25} \caption{\label{model} \myfont \textbf{a}, Illustration of the maximum entropy distribution $P(\Delta t)$ representing the probability of recalling a stimulus $\Delta t$ time steps from the target stimulus (dashed line). In the limit $\beta\rightarrow 0$, the distribution becomes uniform over all past stimuli (left). In the opposite limit $\beta\rightarrow\infty$, the distribution becomes a delta function on the desired stimulus (right). For intermediate amounts of noise, the distribution drops off monotonically (center). \textbf{b}, Resulting internal estimates $\hat{A}$ of the transition structure. For $\beta\rightarrow 0$, the estimates become all-to-all, losing any resemblance to the true structure (left), while for $\beta\rightarrow\infty$, the transition estimates become exact (right). At intermediate precision, the higher-order community structure organically comes into focus (center). \textbf{c-d}, Predictions of the cross-cluster surprisal effect (panel \textbf{c}) and the modular-lattice effect (panel \textbf{d}) as functions of the inverse temperature $\beta$.}}
\end{figure}

To gain an intuition for the model, we consider the infinite-time limit, such that the transition estimates become independent of the particular random walk chosen for analysis. Given a transition matrix $A$, one can show that the asymptotic estimates in our model are equivalent to an average over walks of various lengths, $\hat{A} = \sum_{\Delta t} P(\Delta t) A^{\Delta t + 1}$, which, in turn, can be fashioned into the following analytic expression,
\begin{equation}
\label{analytic}
\hat{A} = (1 - e^{-\beta})A(I - e^{-\beta}A)^{-1},
\end{equation}
where $I$ is the identity matrix (see Methods). The model contains a single free parameter $\beta$, which represents the precision of a person's mental representation. In the limit $\beta\rightarrow\infty$ (no mental errors), our model becomes equivalent to maximum likelihood estimation (Fig. \ref{model}a), and the asymptotic estimates $\hat{A}$ converge to the true transition structure $A$ (Fig \ref{model}b), as expected.\cite{Grimmett-01} Conversely, in the limit $\beta\rightarrow 0$ (overwhelming mental errors), the memory distribution $P(\Delta t)$ becomes uniform across all past nodes (Fig. \ref{model}a), and the mental representation $\hat{A}$ loses all resemblance to the true structure $A$ (Fig. \ref{model}b).

Remarkably, for intermediate values of $\beta$, higher-order features of the transition network, such as communities of densely-connected nodes, come into focus, while some of the fine-scale features, like the edges between communities, fade away (Fig. \ref{model}b). Applying Eq. (\ref{analytic}) to the modular graph, we find that the average expected probability of within-community transitions reaches over 1.6 times the estimated probability of between-community transitions (Fig. \ref{model}c), thus explaining the cross-cluster surprisal effect.\cite{Karuza-02, Kahn-01} Furthermore, we find that the average estimated transition probabilities in the modular graph reach over 1.4 times the estimated probabilities in the lattice graph (Fig. \ref{model}d), thereby predicting the modular-lattice effect. In addition to these higher-order effects, we find that the model also explains previously reported variations in human expectations at the level of individual nodes\cite{Saffran-01, Fiser-01, Kahn-01} (Supplementary Fig. 1). Together, these results demonstrate that the maximum entropy model predicts the qualitative effects of network structure on human reaction times. But can we use the same ideas to quantitatively predict the behavior of particular individuals?

\subsection{Predicting the behavior of individual humans.}

\noindent To model the behavior of individual subjects, we relate the transition estimates in Eqs. (\ref{B}-\ref{Boltzmann}) to predictions about people's reaction times. Given a sequence of nodes $x_1,\hdots, x_{t-1}$, we note that the reaction to the next node $x_t$ is determined by the expected probability of transitioning from $x_{t-1}$ to $x_t$ calculated at time $t-1$, which we denote by $a(t) = \hat{A}_{x_{t-1},x_t}(t-1)$. From this internal anticipation $a(t)$, the simplest possible prediction $\hat{r}(t)$ for a person's reaction time is given by the linear relationship\cite{Neter-01} $\hat{r}(t) = r_0 + r_1a(t)$, where the intercept $r_0$ represents a person's reaction time with zero anticipation and the slope $r_1$ quantifies the strength of the relationship between a person's reactions and their anticipation in our model.\cite{Seber-01}

\begin{figure}
\centering
\includegraphics[width = \textwidth]{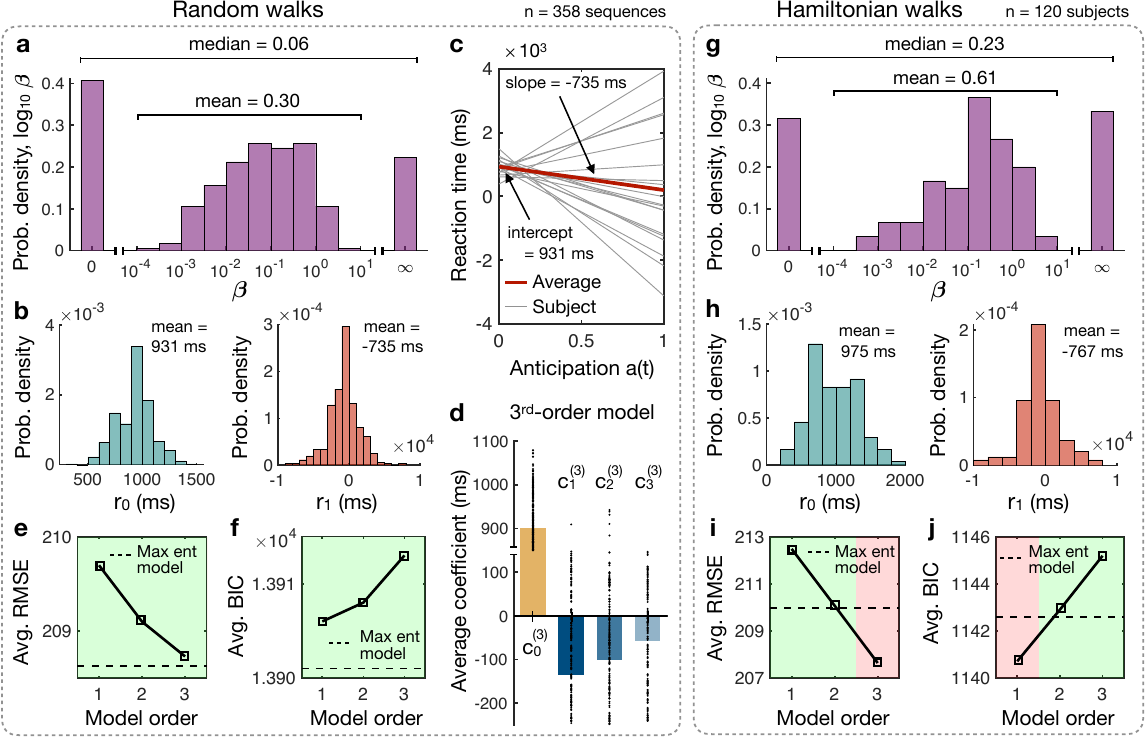} \\
\raggedright
\myfont\textbf{Fig. 4: Predicting reaction times for individual subjects.}
\captionsetup{labelformat=empty}
{\spacing{1.25} \caption{\label{performance} \myfont \textbf{a-f}, Estimated parameters and accuracy analysis for our maximum entropy model across 358 random walk sequences (across 286 subjects; see Methods). \textbf{a}, For the inverse temperature $\beta$, 40 sequences corresponded to the limit $\beta\rightarrow\infty$, 73 corresponded to the limit $\beta\rightarrow 0$. Among the remaining 245 sequences, the average value of $\beta$ was $0.30$. \textbf{b}, Distributions of the intercept $r_0$ (left) and slope $r_1$ (right). \textbf{c}, Predicted reaction time as a function of a subject's internal anticipation. Grey lines indicate 20 randomly-selected sequences, and the red line shows the average prediction over all sequences. \textbf{d}, Linear parameters for the third-order competing model; data points represent individual sequences and bars represent averages. \textbf{e-f}, Comparing the performance of our maximum entropy model with the hierarchy of competing models up to third-order. Root mean squared error (RMSE; \textbf{e}) and Bayesian information criterion (BIC; \textbf{f}) of our model averaged over all sequences (dashed lines) compared to the competing models (solid lines); our model provides the best description of the data across all models considered. \textbf{g-j}, Estimated parameters and accuracy analysis for our maximum entropy model across all Hamiltonian walk sequences (120 subjects). \textbf{g}, For the inverse temperature $\beta$, 20 subjects were best described as performing maximum likelihood estimation $\quad\quad\quad\quad\quad\quad$}}
\end{figure}

To estimate the parameters $\beta$, $r_0$, and $r_1$ that best describe a given individual, we minimize the root mean squared error (RMSE) between their predicted and observed reaction times after regressing out the dependencies on button combination, trial number, and recency (Figs. \ref{experiment}e and \ref{experiment}f; see Methods). The distributions of the estimated parameters are shown in Fig. \ref{performance}a-b for random walks and in Fig. \ref{performance}g-h for Hamiltonian walks. Among the 358 random walk sequences in the modular and lattice graphs (across 286 subjects; see Methods), 40 were best described as performing maximum likelihood estimation ($\beta\rightarrow \infty$) and 73 seemed to lack any notion of the transition structure whatsoever ($\beta\rightarrow 0$), while among the remaining 245 sequences, the average inverse temperature was $\beta = 0.30$. Meanwhile, among the 120 subjects that responded to Hamiltonian walk sequences, 81 appeared to have a non-trivial value of $\beta$, with an average of $\beta = 0.61$. Interestingly, these estimates of $\beta$ roughly correspond to the values for which our model predicts the strongest network effects (Figs. \ref{model}c and \ref{model}d). In the following section, we will compare these values of $\beta$, which are estimated indirectly from people's reaction times, with direct measurements of $\beta$ in an independent memory experiment.

\addtocounter{figure}{-1}
\begin{figure}[t!]
\centering
\raggedright
\captionsetup{labelformat=empty}
{\spacing{1.25} \caption{\myfont ($\beta\rightarrow\infty$), 19 lacked any notion of the transition structure ($\beta\rightarrow 0$), and the remaining 81 subjects had an average value of $\beta = 0.61$. \textbf{h}, Distributions of the intercept $r_0$ (left) and slope $r_1$ (right). \textbf{i}, Average RMSE of our model (dashed line) compared to that of the competing models (solid line); our model maintains higher accuracy than the competing hierarchy up to the second-order model. \textbf{j}, Average BIC of the maximum entropy model (dashed line) compared to that of the competing models (solid line); our model provides a better description of the data than the second- or third-order models. Source data are provided as a Source Data file.}}
\end{figure}

In addition to estimating $\beta$, we also wish to determine whether our model accurately describes individual behavior. Toward this end, we first note that the average slope $r_1$ is large (-735 ms for random walks and -767 ms for Hamiltonian walks), suggesting that the transition estimates in our model $a(t)$ are strongly predictive of human reaction times, and negative, confirming the intuition that increased anticipation yields decreased reaction times (Figs. \ref{performance}b and \ref{performance}h). To examine the accuracy of our model $\hat{r}$, we consider a hierarchy of competing models $\hat{r}^{(\ell)}$, which represent the hypothesis that humans learn explicit representations of the higher-order transition structure. In particular, we denote the $\ell^{\text{th}}$-order transition matrix by $\hat{A}^{(\ell)}_{ij} = n_{ij}^{(\ell)}/\sum_k n_{ik}^{(\ell)}$, where $n_{ij}^{(\ell)}$ counts the number of observed transitions from node $i$ to node $j$ in $\ell$ steps. The model hierarchy takes into account increasingly higher-order transitions, such that the $\ell^{\text{th}}$-order model contains perfect information about transitions up to length $\ell$:
\begin{align}
\label{hierarchy}
\hat{r}^{(0)}(t) &= c^{(0)}_0, \nonumber \\
\hat{r}^{(1)}(t) &= c^{(1)}_0 + c^{(1)}_1a^{(1)}(t), \nonumber \\
&\,\,\, \vdots \nonumber \\
\hat{r}^{(\ell)}(t) &= c^{(\ell)}_0 + \sum_{k = 1}^{\ell} c^{(\ell)}_k a^{(k)}(t),
\end{align}
where $a^{(k)}(t) = \hat{A}^{(k)}_{x_{t-1},x_t}(t-1)$. Each model $\hat{r}^{(\ell)}$ contains $\ell +1$ parameters $c^{(\ell)}_0,\hdots, c^{(\ell)}_{\ell}$, where $c^{(\ell)}_k$ quantifies the predictive power of the $k^{\text{th}}$-order transition structure.

Intuitively, for each model $\hat{r}^{(\ell)}$, we expect $c^{(\ell)}_1, c^{(\ell)}_2,\hdots$ to be negative, reflecting a decrease in reaction times due to increased anticipation, and decreasing in magnitude, such that higher-order transitions are progressively less predictive of people's reaction times. Indeed, considering the third-order model $\hat{r}^{(3)}$ as an example, we find that progressively higher-order transitions are less predictive of human reactions (Fig. \ref{performance}d). However, even the largest coefficient ($c^{(3)}_1 = -135$ ms) is much smaller than the slope in our maximum entropy model ($r_1 = -735$ ms), indicating that the representation $\hat{A}$ is more strongly predictive of people's reaction times than any of the explicit representations $\hat{A}^{(1)}, \hat{A}^{(2)},\hdots$. Indeed, averaging over the random walk sequences, the maximum entropy model achieves higher accuracy than the first three orders of the competing model hierarchy (Fig. \ref{performance}e) -- this is despite the fact that the third-order model even contains one more parameter. To account for differences in the number of parameters, we additionally compare the average Bayesian information criterion (BIC) of our model with that of the competing models, finding that the maximum entropy model provides the best description of the data (Fig. \ref{performance}f).

Similarly, averaging over the Hamiltonian walk sequences, the maximum entropy model provides more accurate predictions than the first two competing models (Fig. \ref{performance}i) and provides a lower BIC than the second and third competing models (Fig. \ref{performance}j). Notably, even in Hamiltonian walks, the maximum entropy model provides a better description of human reaction times than the second-order competing model, which has the same number of parameters. However, we remark that the first-order competing model has a lower BIC than the maximum entropy model (Fig. \ref{performance}j), suggesting that humans may focus on first-order rather than higher-order statistics during Hamiltonian walks -- an interesting direction for future research. On the whole, these results indicate that the free energy principle, and the resulting maximum entropy model, are consistently more effective at describing human reactions than the hypothesis that people learn explicit representations of the higher-order transition structure.

\subsection{Directly probing the memory distribution.}

\noindent Throughout our discussion, we have argued that errors in memory shape human representations in predictable ways, a perspective that has received increasing attention in recent years.\cite{Richards-01, Collins-01, Collins-02} While our framework explains specific aspects of human behavior, there exist alternative perspectives that might yield similar predictions. For example, one could imagine a Bayesian learner with a non-Markov prior that ``integrates" the transition structure over time, even without sustaining errors in memory or learning. Additionally, Eq. (\ref{analytic}) resembles the successor representation in reinforcement learning,\cite{Dayan-01, Gershman-01} which assumes that, rather than shuffling the order of past stimuli, humans are instead planning their responses multiple steps in advance (see Supplementary Discussion). In order to distinguish our framework from these alternatives, here we provide direct evidence for precisely the types of mental errors predicted by our model.

In the construction and testing of our model, we have developed a series of predictions concerning the shape of the memory distribution $P(\Delta t)$, which, to recall, represents the probability of remembering the stimulus at time $t - \Delta t$ instead of the target stimulus at time $t$. We first assumed that $P(\Delta t)$ decreases monotonically. Second, to make quantitative predictions, we employed the free energy principle, leading to the prediction that $P$ drops off exponentially quickly with $\Delta t$ (Eq. (\ref{Boltzmann})). Finally, when fitting the model to individual subjects, we estimated an average inverse temperature $\beta$ between $0.30$ for random walks and $0.61$ for Hamiltonian walks.

\begin{figure}[t]
\centering
\includegraphics[width = .8\textwidth]{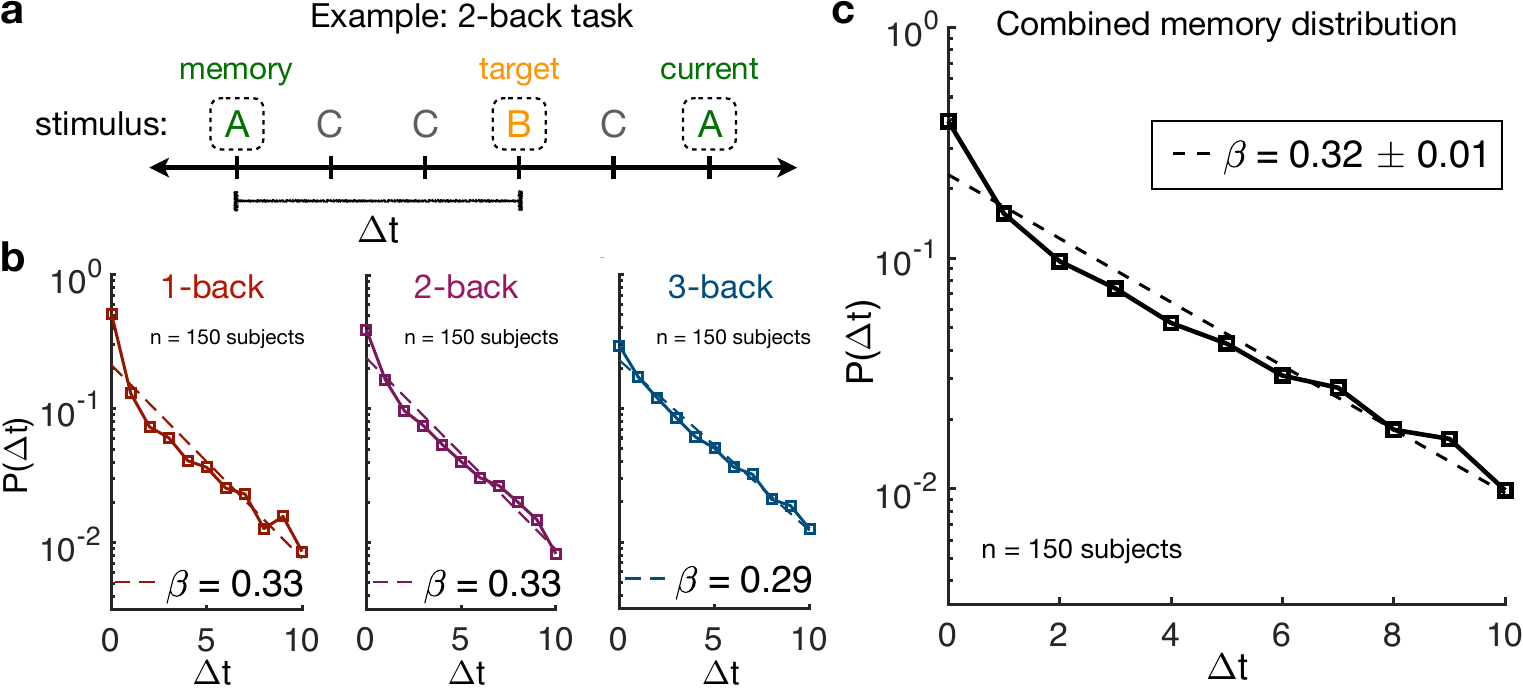} \\
\raggedright
\myfont\textbf{Fig. 5: Measuring the memory distribution in an $n$-back experiment.}
\captionsetup{labelformat=empty}
{\spacing{1.25} \caption{\label{nback} \myfont \textbf{a}, Example of the 2-back memory task. Subjects view a sequence of stimuli (letters) and respond to each stimulus indicating whether it matches the target stimulus from two trials before. For each positive response that the current stimulus matches the target, we measure $\Delta t$ by calculating the number of trials between the last instance of the current stimulus and the target. \textbf{b}, Histograms of $\Delta t$ (i.e., measurements of the memory distribution $P(\Delta t)$) across all subjects in the 1-, 2-, and 3-back tasks. Dashed lines indicate exponential fits to the observed distributions. The inverse temperature $\beta$ is estimated for each task to be the negative slope of the exponential fit. \textbf{c}, Memory distribution aggregated across the three $n$-back tasks. Dashed line indicates an exponential fit. We report a combined estimate of the inverse temperature $\beta = 0.32 \pm 0.01$, where the standard deviation is estimated from 1,000 bootstrap samples of the combined data. Measurements were on independent subjects. Source data are provided as a Source Data file.}}
\end{figure}

To test these three predictions directly, we conducted a standard $n$-back memory experiment. Specifically, we presented subjects with sequences of letters on a screen, and they were asked to respond to each letter indicating whether or not it was the same as the letter that occurred $n$ steps previously; for each subject, this process was repeated for the three conditions $n=1,2,$ and $3$. To measure the memory distribution $P(\Delta t)$, we considered all trials on which a subject responded positively that the current stimulus matched the target. For each such trial, we looked back to the last time that the subject did in fact observe the current stimulus and we recorded the distance (in trials) between this observation and the target (Fig. \ref{nback}a). In this way, we were able to treat each positive response as a sample from the memory distribution $P(\Delta t)$.

The measurements of $P$ for the 1-, 2-, and 3-back tasks are shown in Figure \ref{nback}b, and the combined measurement of $P$ across all conditions is shown in Figure \ref{nback}c. Notably, the distributions decrease monotonically and maintain consistent exponential forms, even out to $\Delta t = 10$ trials from the target stimulus, thereby providing direct evidence for the Boltzmann distribution (Eq. (\ref{Boltzmann})). Moreover, fitting an exponential curve to each distribution, we can directly estimate the inverse temperature $\beta$. Remarkably, the value $\beta = 0.32 \pm 0.1$ estimated from the combined distribution (Fig. \ref{nback}c) falls within the range of values estimated from our reaction time experiments (Figs. \ref{performance}a and \ref{performance}g), nearly matching the average value $\beta = 0.30$ for random walk sequences (Fig. \ref{performance}a).

To further strengthen the link between mental errors and people's internal representations, we then asked subjects to perform the original serial response task (Fig. \ref{experiment}), and for each subject, we estimated $\beta$ using the two methods described above: (i) directly measuring $\beta$ in the $n$-back experiment, and (ii) indirectly estimating $\beta$ in the serial response experiment. Comparing these two estimates across subjects, we find that they are significantly related with Spearman correlation $r_s = 0.28$ ($p = 0.047$, permutation test), while noting that we do not use the Pearson correlation because $\beta$ is not normally distributed (Anderson-Darling test,\cite{Stephens-01} $p < 0.001$ for the serial response task and $p = 0.013$ for the $n$-back task). Together, these results demonstrate not only the existence of the particular form of mental errors predicted by our model -- down to the specific value of $\beta$ -- but also the relationship between these mental errors and people's internal estimates of the transition structure.

\subsection{Network structure guides reactions to novel transitions.}

\noindent Given a model of human behavior, it is ultimately interesting to make testable predictions. Thus far, in keeping with the majority of existing research,\cite{Schapiro-01, Karuza-01, Karuza-02, Kahn-01, Saffran-01, Fiser-01} we have focused on static transition graphs, wherein the probability $A_{ij}$ of transitioning from state $i$ to state $j$ remains constant over time. However, the statistical structures governing human life are continually shifting,\cite{Whitehead-01, Wang-01} and people are often forced to respond to rare or novel transitions.\cite{Wolfe-01, Tria-01} Here we show that, when confronted with a novel transition -- or a \emph{violation} of the preexisting transition network -- not only are people surprised, but the magnitude of their surprise depends critically on the topology of the underlying network.

We consider a ring graph where each node is connected to its nearest and next-nearest neighbors (Fig. \ref{violations}a). We asked subjects to respond to sequences of 1500 nodes drawn as random walks on the ring graph, but with 50 violations randomly interspersed. These violations were divided into two categories: short violations of topological distance two and long violations of topological distances three and four (Fig. \ref{violations}a). Using maximum likelihood estimation (Eq. (\ref{MLE})) as a guide, one would na\"{i}vely expect people to be equally surprised by all violations -- indeed, each violation has never been seen before. In contrast, our model predicts that that surprise should depend crucially on the topological distance of a violation in the underlying graph, with topologically longer violations inducing increased surprise over short violations (Fig. \ref{violations}b).

\begin{figure}[t]
\centering
\includegraphics[width = \textwidth]{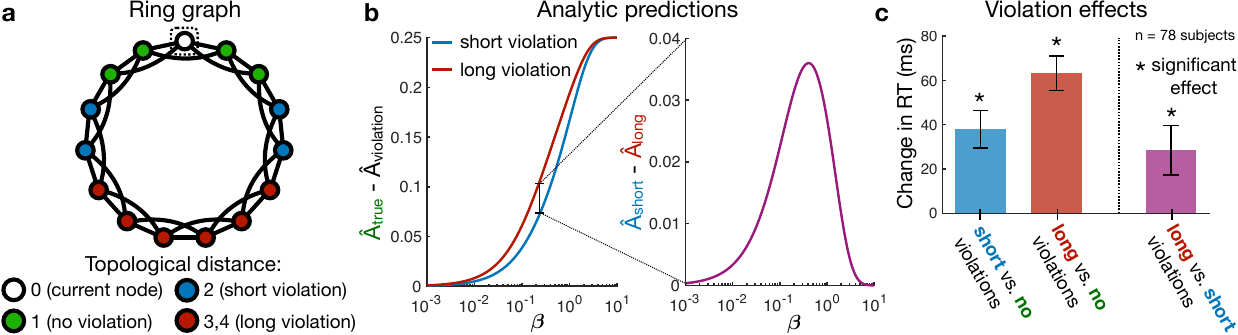} \\
\raggedright
\myfont\textbf{Fig. 6: Network violations yield surprise that grows with topological distance.}
\captionsetup{labelformat=empty}
{\spacing{1.25} \caption{\label{violations} \myfont \textbf{a}, Ring graph consisting of 15 nodes, where each node is connected to its nearest neighbors and next-nearest neighbors on the ring. Starting from the boxed node, a sequence can undergo a standard transition (green), a short violation of the transition structure (blue), or a long violation (red). \textbf{b}, Our model predicts that subjects' anticipations of both short (blue) and long (red) violations should be weaker than their anticipations of standard transitions (left). Furthermore, we predict that subjects' anticipations of violations should decrease with increasing topological distance (right). \textbf{c}, Average effects of network violations across 78 subjects, estimated using a mixed effects model (see Supplementary Tables 10 and 11), with error bars indicating one standard deviation from the mean. We find that standard transitions yield quicker reactions than both short violations ($p < 0.001$, $t = 4.50$, $\text{df} = 7.15\times 10^4$) and long violations ($p < 0.001$, $t = 8.07$, $\text{df} = 7.15\times 10^4$). Moreover, topologically shorter violations induce faster reactions than long violations ($p = 0.011$, $t = 2.54$, $\text{df} = 3.44\times 10^3$), thus confirming the predictions of our model. Measurements were on independent subjects, and statistical significance was computed using two-sided $F$-tests. Source data are provided as a Source Data file.}}
\end{figure}

In the data, we find that all violations give rise to sharp increases in reaction times relative to standard transitions (Fig. \ref{violations}c; Supplementary Table 10), indicating that people are in fact learning the underlying transition structure. Moreover, we find that reaction times for long violations are 28 ms longer than those for short violations ($p = 0.011$, $F$-test; Fig. \ref{violations}c; Supplementary Table 11). Additionally, we confirm that the effects of network violations are not simply driven by stimulus recency (Supplementary Figs. 4 and 5). These observations suggest that people learn the topological distances between all nodes in the transition graph, not just those pairs for which a transition has already been observed.\cite{Whitehead-01, Wolfe-01, Wang-01, Tria-01}

\section*{Discussion}

Daily life is filled with sequences of items that obey an underlying network architecture, from networks of word and note transitions in natural language and music to networks of abstract relationships in classroom lectures and literature.\cite{Bousfield-01, Fiser-01, Friederici-01, Tompson-01, Reynolds-01} How humans infer and internally represent these complex structures are questions of fundamental interest.\cite{Meyniel-01, Dehaene-01, Piantadosi-01, Tenenbaum-01} Recent experiments in statistical learning have established that human representations depend critically on the higher-order organization of probabilistic transitions, yet the underlying mechanisms remain poorly understood.\cite{Schapiro-01, Karuza-01, Karuza-03, Kahn-01}

Here we show that network effects on human behavior can be understood as stemming from mental errors in people's estimates of the transition structure, while noting that future work should focus on disambiguating the role of recency.\cite{Murdock-01, Baddeley-01} We use the free energy principle to develop a model of human expectations that explicitly accounts for the brain's natural tendency to minimize computational complexity -- that is, to maximize entropy.\cite{Ortega-01, Ortega-02, Friston-01} Indeed, the brain must balance the benefits of making accurate predictions against the computational costs associated with such predictions.\cite{Tversky-01, De-01, Vinje-01, Tononi-01, Cohen-01, Wickelgren-01, Gershman-02} This competition between accuracy and efficiency induces errors in people's internal representations, which, in turn, explains with notable accuracy an array of higher-order network phenomena observed in human experiments.\cite{Schapiro-01, Karuza-01, Karuza-03, Kahn-01} Importantly, our model admits a concise analytic form (Eq. (\ref{analytic})) and can be used to predict human behavior on a person-by-person basis (Fig. \ref{performance}).

This work inspires directions for future research, particularly with regard to the study and design of optimally learnable network structures. Given the notion that densely connected communities help to mitigate the effects of mental errors on people's internal representations, we anticipate that networks with high ``learnability" will possess a hierarchical community structure.\cite{Arenas-01} Interestingly, such hierarchical organization has already been observed in a diverse range of real world networks, from knowledge and language graphs\cite{Guimera-01} to social networks and the World Wide Web.\cite{Ravasz-01} Could it be that these networks have evolved so as to facilitate accurate representations in the minds of the humans using and observing them? Questions such as this demonstrate the importance of having simple principled models of human representations and point to the promising future of this research endeavor.

\begin{methods}

\noindent \textbf{Maximum entropy model and the infinite-sequence limit.} Here we provide a more thorough derivation of our maximum entropy model of human expectations, with the goal of fostering intuition. Given a matrix of erroneous transition counts $\tilde{n}_{ij}$, our estimate of the transition structure is given by $\hat{A}_{ij} = \tilde{n}_{ij}/\sum_k \tilde{n}_{ik}$. When observing a sequence of nodes $x_1,x_2,\hdots$, in order to construct the counts $\tilde{n}_{ij}$, we assume that humans use the following recursive rule: $\tilde{n}_{ij}(t+1) = \tilde{n}_{ij}(t) + B_t(i)\left[j = x_{t+1}\right]$, where $B_t(i)$ denotes the belief, or perceived probability, that node $i$ occurred at the previous time $t$. This belief, in turn, can be written in terms of the probability $P(\Delta t)$ of accidentally recalling the node that occurred $\Delta t$ time steps from the desired node at time $t$: $B_t(i) = \sum_{\Delta t = 0}^{t-1} P(\Delta t) \left[i = x_{t-\Delta t}\right]$.

In order to make quantitative predictions about people's estimates of a transition structure, we must choose a mathematical form for $P(\Delta t)$. To do so, we leverage the free energy principle\cite{Ortega-01}: When building mental models, the brain is finely-tuned to simultaneously minimize errors and computational complexity. The average error associated with a candidate distribution $Q(\Delta t)$ is assumed to be the average distance in time of the recalled node from the target node, denoted $E(Q) = \sum_{\Delta t} Q(\Delta t)\Delta t$. Furthermore, Shannon famously proved that the only suitable choice for the computational cost of a candidate distribution is its negative entropy,\cite{Shannon-01} denoted $-S(Q) = \sum_{\Delta t} Q(\Delta t)\log Q(\Delta t)$. Taken together, the total cost associated with a distribution $Q(\Delta t)$ is given by the free energy $F(Q) = \beta E(Q) - S(Q)$, where $\beta$, referred to as the inverse temperature, parameterizes the relative importance of minimizing errors versus computational costs. By minimizing $F$ with respect to $Q$, we arrive at the Boltzmann distribution $P(\Delta t) = e^{-\beta \Delta t}/Z$, where $Z$ is the normalizing partition function.\cite{Jaynes-01} We emphasize that this mathematical form for $P(\Delta t)$ followed directly from our free energy assumption about resource constraints in the brain.

To gain an analytic intuition for the model without referring to a particular random walk, we consider the limit of an infinitely long sequence of nodes. To begin, we consider a sequence $x_1,\hdots, x_T$ of length $T$. At the end of this sequence, the counting matrix takes the form
\begin{align}
\label{counts}
\tilde{n}_{ij}(T) &= \sum_{t = 1}^{T-1} B_t(i)\left[j = x_{t+1}\right] \nonumber \\
&= \sum_{t = 1}^{T-1} \Bigg(\sum_{\Delta t = 0}^{t-1}P(\Delta t)\left[i = x_{t-\Delta t}\right]\Bigg)\left[j = x_{t+1}\right].
\end{align}
Dividing both sides by $T$, the right-hand side becomes a time average, which by the ergodic theorem converges to an expectation over the transition structure in the limit $T\rightarrow\infty$,
\begin{equation}
\label{limit}
\lim_{T\rightarrow \infty} \frac{\tilde{n}_{ij}(T)}{T} = \sum_{\Delta t = 0}^{\infty} P(\Delta t) \left<\left[i = x_{t-\Delta t}\right]\left[j = x_{t+1}\right]\right>_A,
\end{equation}
where $\left<\cdot\right>_A$ denotes an expectation over random walks in $A$. We note that the expectation of an identity function is simply a probability, such that $\left<\left[i = x_{t-\Delta t}\right]\left[j = x_{t+1}\right]\right>_A = p_i \left(A^{\Delta t + 1}\right)_{ij}$, where $p_i$ is the long-run probability of node $i$ appearing in the sequence and $\left(A^{\Delta t + 1}\right)_{ij}$ is the probability of randomly walking from node $i$ to node $j$ in $\Delta t + 1$ steps. Putting these pieces together, we find that the expectation $\hat{A}$ converges to a concise mathematical form,
\begin{align}
\lim_{T\rightarrow\infty} \hat{A}_{ij}(T) &= \lim_{T\rightarrow\infty} \frac{\tilde{n}_{ij}(T)}{\sum_k \tilde{n}_{ik}(T)} \nonumber \\
&= \frac{p_i \sum_{\Delta t = 0}^{\infty} P(\Delta t) \left(A^{\Delta t + 1}\right)_{ij}}{p_i} \nonumber \\
&= \sum_{\Delta t = 0}^{\infty} P(\Delta t) \left(A^{\Delta t + 1}\right)_{ij}.
\end{align}
Thus far, we have not appealed to our maximum entropy form for $P(\Delta t)$. It turns out that doing so allows us to write down an analytic expression for the long-time expectations $\hat{A}$ simply in terms of the transition structure $A$ and the inverse temperature $\beta$. Noting that $Z = \sum_{\Delta t = 0}^{\infty} e^{-\beta \Delta t} = 1/(1- e^{-\beta})$ and $\sum_{\Delta t = 0}^{\infty} \left( e^{-\beta} A\right)^{\Delta t} = \left(I - e^{-\beta}A\right)^{-1}$, we have
\begin{align}
\hat{A} &= \sum_{\Delta t = 0}^{\infty} P(\Delta t) A^{\Delta t + 1} \nonumber \\
&= \frac{1}{Z}A\sum_{\Delta t = 0}^{\infty} \left(e^{-\beta} A\right)^{\Delta t} \nonumber \\
&= \left(1 - e^{-\beta}\right)A\left(I - e^{-\beta}A\right)^{-1}.
\end{align}
This simple formula for the representation $\hat{A}$ is the basis for all of our analytic predictions (Figs. \ref{model}c, \ref{model}d, and \ref{violations}b) and is closely related to notions of communicability in complex network theory.\cite{Estrada-01,Estrada-02}

\noindent \textbf{Experimental setup for serial response tasks.} Subjects performed a self-paced serial reaction time task using a computer screen and keyboard. Each stimulus was presented as a horizontal row of five grey squares; all five squares were shown at all times. The squares corresponded spatially with the keys `Space', `H', `J', `K', and `L', with the left square representing `Space' and the right square representing `L' (Fig. \ref{experiment}b). To indicate a target key or pair of keys for the subject to press, the corresponding squares would become outlined in red (Fig. \ref{experiment}a). When subjects pressed the correct key combination, the squares on the screen would immediately display the next target. If an incorrect key or pair of keys was pressed, the message `Error!' was displayed on the screen below the stimuli and remained until the subject pressed the correct key(s). The order in which stimuli were presented to each subject was prescribed by either a random walk or a Hamiltonian walk on a graph of $N=15$ nodes, and each sequence consisted of 1500 stimuli. For each subject, one of the 15 key combinations was randomly assigned to each node in the graph (Fig. \ref{experiment}a). Across all graphs, each node was connected to its four neighboring nodes with a uniform $0.25$ transition probability. Importantly, given the uniform edge weights and homogeneous node degrees ($k=4$), the only differences between the transition graphs lay in their higher-order structure. 

In the first experiment, we presented subjects with random walk sequences drawn from two different graph topologies: a \textit{modular} graph with three communities of five densely-connected nodes and a \textit{lattice} graph representing a $3\times 5$ grid with periodic boundary conditions (Fig. \ref{experiment}c). The purpose of this experiment was to demonstrate the systematic dependencies of human reaction times on higher-order network structure, following similar results reported in recent literature.\cite{Kahn-01, Karuza-02} In particular, we demonstrate two higher-order network effects: In the \textit{cross-cluster surprisal} effect, average reaction times for within-cluster transitions in the modular graph are significantly faster than reaction times for between-cluster transitions (Fig. \ref{effects}a); and in the \textit{modular-lattice} effect, average reaction times in the modular graph are significantly faster than reaction times in the lattice graph (Fig. \ref{effects}b).

In the second experiment, we presented subjects with Hamiltonian walk sequences drawn from the modular graph. Specifically, each sequence consisted of 700 random walk trials (intended to allow each subject to learn the graph structure), followed by eight repeats of 85 random walk trials and 15 Hamiltonian walk trials (see Supplementary Discussion).\cite{Schapiro-01} Importantly, we find that the cross-cluster surprisal effect remains significant within the Hamiltonian walk trials (Fig. \ref{effects}a).

In the third experiment, we considered a \textit{ring} graph where each node was connected to its nearest and next-nearest neighbors in the ring (Fig. \ref{violations}a). In order to study the dependence of human expectations on violations to the network structure, the first 500 trials for each subject constituted a standard random walk, allowing each subject time to develop expectations about the underlying transition structure. Across the final 1000 trials, we randomly distributed 50 network violations: 20 short violations of topological distance two and 30 long violations, 20 of topological distance three and 10 of topological distance four (Fig. \ref{violations}a). As predicted by our model, we found a novel \textit{violations} effect, wherein violations of longer topological distance give rise to larger increases in reaction times than short, local violations (Figs. \ref{violations}b and \ref{violations}c).

\noindent \textbf{Data analysis for serial response tasks.} To make inferences about subjects' internal expectations based on their reaction times, we used more stringent filtering techniques than previous experiments when pre-processing the data.\cite{Kahn-01} Across all experiments, we first excluded from analysis the first 500 trials, in which subjects' reaction times varied wildly (Fig. \ref{experiment}e), focusing instead on the final 1000 trials (or simply on the Hamiltonian trials in the second experiment), at which point subjects had already developed internal expectations about the transition structures. We then excluded all trials in which subjects responded incorrectly. Finally, we excluded reaction times that were implausible, either three standard deviations from a subject's mean reaction time or below 100 ms. Furthermore, when measuring the network effects in all three experiments (Figs. \ref{effects} and \ref{violations}), we also excluded reaction times over 3500 ms for implausibility. When estimating the parameters of our model and measuring model performance in the first two experiments (Fig. \ref{performance}), to avoid large fluctuations in the results based on outlier reactions, we were even more stringent, excluding all reaction times over 2000 ms. Taken together, when measuring the cross-cluster surprisal and modular-lattice effects (Fig. \ref{effects}), we used an average of 931 trials per subject; when estimating and evaluating our model (Fig. \ref{performance}), we used an average of 911 trials per subject; and when measuring the violation effects (Fig. \ref{violations}), we used an average of 917 trials per subject. To ensure that our results are robust to particular choices in the data processing, we additionally studied all 1500 trials for each subject rather than just the final 1000, confirming that both the cross-cluster surprisal and modular-lattice effects remain significant across all trials (Supplementary Tables 6 and 7).

\noindent \textbf{Measurement of network effects using mixed effects models.} In order to extract the effects of higher-order network structure on subjects' reaction times, we used linear mixed effects models, which have become prominent in human research where many measurements are made for each subject.\cite{Schall-01,Baayen-01} Put simply, mixed effects models generalize standard linear regression techniques to include both \textit{fixed} effects, which are constant across subjects, and \textit{random} effects, which vary between subjects. Compared with standard linear models, mixed effects models allow for differentiation between effects that are subject-specific and those that persist across an entire population. Here, all models were fit using the \texttt{fitlme} function in MATLAB (R2018a), and random effects were chosen as the maximal structure that (i) allowed model convergence and (ii) did not include effects whose 95\% confidence intervals overlapped with zero.\cite{Hox-01} In what follows, when defining mixed effects models, we employ the standard R notation.\cite{Bates-01}

First, we considered the cross-cluster surprisal effect (Fig. \ref{effects}a). Since we were only interested in measuring higher-order effects of the network topology on human reaction times, it was important to regress out simple biomechanical dependencies on the target button combinations (Fig. \ref{experiment}d), the natural quickening of reactions with time (Fig. \ref{experiment}e), and the effects of recency on reaction times.\cite{Murdock-01, Baddeley-01} Also, for the first experiment, since some subjects responded to both the modular and lattice graphs (see Experimental Procedures), it was important to account for changes in reaction times due to which stage of the experiment a subject was in. To measure the cross-cluster surprisal effect, we fit a mixed effects model with the formula `$\text{RT}\sim \log(\text{Trial})*\text{Stage} + \text{Target} + \text{Recency} + \text{Trans}\_\text{Type} + (1 + \log(\text{Trial})*\text{Stage} + \text{Recency} + \text{Trans}\_\text{Type} \,|\, \text{ID})$', where RT is the reaction time, Trial is the trial number (we found that $\log(\text{Trial})$ was far more predictive of subjects' reaction times than the trial number itself), Stage is the stage of the experiment (either one or two), Target is the target button combination, Recency is the number of trials since the last instance of the current stimulus, Trans$\_$Type is the type of transition (either within-cluster or between-cluster), and ID is each subject's unique ID. Fitting this mixed effects model to the random walk data in the first experiment (Supplementary Table 1), we found a 35 ms increase in reaction times ($p < 0.001$, $F$-test) for between-cluster transitions relative to within-cluster transitions (Fig. \ref{effects}a). Similarly, fitting the same mixed effects model but without the variable $Stage$ to the Hamiltonian walk data in the second experiment (Supplementary Table 4), we found a 36 ms increase in reaction times ($p < 0.001$, $F$-test) for between- versus within-cluster transitions (Fig. \ref{effects}a). We note that because reaction times are not Gaussian distributed, it is fairly standard to perform a log transformation. However, for the above result as well as those that follow, we find the same qualitative effects with or without a log transformation.

Second, we studied the modular-lattice effect (Fig. \ref{effects}b). To do so, we fit a mixed effects model with the formula `$\text{RT}\sim \log(\text{Trial})*\text{Stage} + \text{Target} + \text{Recency} + \text{Graph} + (1 + \log(\text{Trial})*\text{Stage} + \text{Recency} + \text{Graph} \,|\, \text{ID})$', where Graph represents the type of transition network, either modular or lattice. Fitting this mixed effects model to the data in the first experiment (Supplementary Table 2), we found a fixed 23 ms increase in reaction times ($p < 0.001$, $F$-test) in the lattice graph relative to the modular graph (Fig. \ref{effects}b).

Finally, we considered the effects of violations of varying topological distance in the ring lattice (Fig. \ref{violations}c). We fit a mixed effects model with the formula `$\text{RT}\sim \log(\text{Trial}) + \text{Target} + \text{Recency} + \text{Top}\_\text{Dist} + (1 + \log(\text{Trial}) + \text{Recency} + \text{Top}\_\text{Dist} \,|\, \text{ID})$', where Top\_Dist represents the topological distance of a transition, either one for a standard transition, two for a short violation, or three for a long violation. Fitting the model to the data in the third experiment (Supplementary Tables 10 and 11), we found a 38 ms increase in reaction times for short violations relative to standard transitions ($p < 0.001$, $F$-test), a 63 ms increase in reaction times for long violations relative to standard transitions ($p < 0.001$, $F$-test), and a 28 ms increase in reaction times for long violations relative to short violations ($p = 0.011$, $F$-test). Put simply, people are more surprised by violations to the network structure that take them further from their current position in the network, suggesting that people have an implicit understanding of the topological distances between nodes in the network.

\noindent \textbf{Estimating parameters and making quantitative predictions.} Given an observed sequence of nodes $x_1,\hdots,x_{t-1}$, and given an inverse temperature $\beta$, our model predicts the anticipation, or expectation, of the subsequent node $x_t$ to be $a(t) = \hat{A}_{x_{t-1},x_t}(t-1)$. In order to quantitatively describe the reactions of an individual subject, we must relate the expectations $a(t)$ to predictions about a person's reaction times $\hat{r}(t)$ and then calculate the model parameters that best fit the reactions of an individual subject. The simplest possible prediction is given by the linear relation $\hat{r}(t) = r_0 + r_1a(t)$, where the intercept $r_0$ represents a person's reaction time with zero anticipation and the slope $r_1$ quantifies the strength with which a person's reaction times depend on their internal expectations.

In total, our predictions $\hat{r}(t)$ contain three parameters ($\beta$, $r_0$, and $r_1$), which must be estimated from the reaction time data for each subject. Before estimating these parameters, however, we first regress out the dependencies of each subject's reaction times on the button combinations, trial number, and recency using a mixed effects model of the form `$\text{RT}\sim \log(\text{Trial})*\text{Stage} + \text{Target} + \text{Recency} + (1 + \log(\text{Trial})*\text{Stage} + \text{Recency} \,|\, \text{ID})$', where all variables were defined in the previous section. Then, to estimate the model parameters that best describe an individual's reactions, we minimize the RMS prediction error with respect to each subject's observed reaction times, $\text{RMSE} = \sqrt{\frac{1}{T}\sum_t(r(t) - \hat{r}(t))^2}$, where $T$ is the number of trials. We note that, given a choice for the inverse temperature $\beta$, the linear parameters $r_0$ and $r_1$ can be calculated analytically using standard linear regression techniques. Thus, the problem of estimating the model parameters can be restated as a one-dimensional minimization problem; that is, minimizing RMSE with respect to the inverse temperature $\beta$. To find the global minimum, we began by calculating RMSE along 100 logarithmically-spaced values for $\beta$ between $10^{-4}$ and $10$. Then, starting at the minimum value of this search, we performed gradient descent until the gradient fell below an absolute value of $10^{-6}$. For a derivation of the gradient of the RMSE with respect to the inverse temperature $\beta$, we point the reader to the Supplementary Discussion. Finally, in addition to the gradient descent procedure described above, for each subject we also manually checked the RMSE associated with the two limits $\beta\rightarrow 0$ and $\beta\rightarrow\infty$. The resulting model parameters are shown in Figs. \ref{performance}a and \ref{performance}b for random walk sequences and Figs. \ref{performance}g and \ref{performance}h for Hamiltonian walk sequences.

\noindent \textbf{Experimental setup for $n$-back memory task.} Subjects performed a series of $n$-back memory tasks using a computer screen and keyboard. Each subject observed a random sequence of the letters `B', 'D', 'G', 'T', and 'V', wherein each letter was randomly displayed in either upper or lower case. The subjects responded on each trial using the keyboard to indicate whether or not the current letter was the same as the letter that occurred $n$ trials previously. For each subject, this task was repeated for the conditions $n = 1,2,$ and $3$, and each condition consisted of a sequence of 100 letters. The three conditions were presented in a random order to each subject. After the $n$-back task, each subject then performed a serial response task (equivalent to the first experiment described above) consisting of 1500 random walk trials drawn from the modular graph.

\noindent \textbf{Data analysis for $n$-back memory task.} In order to estimate the inverse temperature $\beta$ for each subject from their $n$-back data, we directly measured their memory distribution $P(\Delta t)$. As described in the main text, we treated each positive response indicating that the current stimulus matched the target stimulus as a sample of $P(\Delta t)$ by measuring the distance in trials $\Delta t$ between the last instance of the current stimulus and the target (Fig. \ref{nback}a). For each subject, we combined all such samples across the three conditions $n=1,2,$ and $3$ to arrive at a histogram for $\Delta t$. In order to generate robust estimates for the inverse temperature $\beta$, we generated 1000 bootstrap samples of the $\Delta t$ histogram for each subject. For each sample, we calculated a linear fit to the distribution $P(\Delta t)$ on log-linear axes within the domain $0 \le \Delta t \le 4$ (note that we could not carry the fit out to $\Delta t = 10$ because the data is much sparser for individual subjects). To ensure that the logarithm of $P(\Delta t)$ was well defined for each sample -- that is, to ensure that $P(\Delta t) > 0$ for all $\Delta t$ -- we added one count to each value of $\Delta t$. We then estimated the inverse temperature $\beta$ for each sample by calculating the negative slope of the linear fit of $\log P(\Delta t)$ versus $\Delta t$. To arrive at an average estimate of $\beta$ for each subject, we averaged $\beta$ across the 1000 bootstrap samples. Finally, we compared these estimates of $\beta$ from the $n$-back experiment with estimates of $\beta$ from subjects' reaction times in the subsequent serial response task, as described above. We found that these two independent estimates of people's inverse temperatures are significantly correlated (excluding subjects for which $\beta = 0$ or $\beta \rightarrow \infty$), with a Spearman coefficient $r_s = 0.28$ ($p = 0.047$, permutation test). We note that we do not use the Pearson correlation coefficient because the estimates for $\beta$ are not normally distributed for either the reaction time task ($p < 0.001$) nor the $n$-back task ($p = 0.013$) according to the Anderson-Darling test.\cite{Stephens-01} This non-normality can be clearly seen in the distributions of $\beta$ in Figs. \ref{performance}a and \ref{performance}g.

\noindent \textbf{Experimental procedures.} All participants provided informed consent in writing and experimental methods were approved by the Institutional Review Board of the University of Pennsylvania. In total, we recruited 634 unique participants to complete our studies on Amazon's Mechanical Turk. For the first serial response experiment, 101 participants only responded to sequences drawn from the modular graph, 113 participants only responded to sequences drawn from the lattice graph, and 72 participants responded to sequences drawn from both the modular and lattice graphs in back-to-back (counter-balanced) sessions for a total of 173 exposures to the modular graph and 185 exposures to the lattice graph. For the second experiment, we recruited 120 subjects to respond to random walk sequences with Hamiltonian walks interspersed, as described in the Supplementary Discussion. For the third experiment, we recruited 78 participants to respond to sequences drawn from the ring graph with violations randomly interspersed. For the $n$-back experiment, 150 subjects performed the $n$-back task and, of those, 88 completed the subsequent serial response task. Worker IDs were used to exclude duplicate participants between experiments, and all participants were financially remunerated for their time. In the first experiment, subjects were paid up to \$11 for up to an estimated 60 minutes: \$3 per network for up to two networks, \$2 per network for correctly responding on at least 90\% of the trials, and \$1 for completing the entire task. In the second and third experiments, subjects were paid up to \$7.50 for an estimated 30 minutes: \$5.50 for completing the experiment and \$2 for correctly responding on at least 90\% of the trials. In the $n$-back experiment, subjects were paid up to \$8.50 for an estimated 45 minutes: \$7 for completing the entire experiment and \$1.50 for correctly responding on at least 90\% of the serial response trials.

At the beginning of each experiment, subjects were provided with the following instructions: ``In a few minutes, you will see five squares shown on the screen, which will light up as the experiment progresses. These squares correspond with keys on your keyboard, and your job is to watch the squares and press the corresponding key when that square lights up." For the 72 subjects that responded to both the modular and lattice graphs in the first experiment, an additional piece of information was also provided: ``This part will take around 30 minutes, followed by a similar task which will take another 30 minutes.'' Before each experiment began, subjects were given a short quiz to verify that they had read and understood the instructions. If any questions were answered incorrectly, subjects were shown the instructions again and asked to repeat the quiz until they answered all questions correctly. Next, all subjects were shown a 10-trial segment that did not count towards their performance; this segment also displayed text on the screen explicitly telling the subject which keys to press on their keyboard. Subjects then began their 1500-trial experiment. For the subjects that responded to both the modular and lattice graphs, a brief reminder was presented before the second graph, but no new instructions were given. After completing each experiment, subjects were presented with performance information and their bonus earned, as well as the option to provide feedback.

\end{methods}

\section*{Data Availability}

Source data for Fig. \ref{experiment} are provided in Supplementary Data File 1. Source data for Fig. \ref{effects}, Supplementary Figs. 2 and 3, and Supplementary Tables 1-9 are provided in Supplementary Data File 2. Source data for Fig. \ref{nback} are provided in Supplementary Data File 3. Source data for Fig. \ref{violations}, Supplementary Figs. 4 and 5, and Supplementary Tables 10 and 11 are provided in Supplementary Data File 4. Source data from Supplementary Fig. 1 are provided in Supplementary Data File 5.

\section*{Code Availability}

The code that supports the findings of this study is available from the corresponding author upon reasonable request.





\newpage

\begin{addendum}
\item[Acknowledgements] We thank Pedro Ortega for discussions. D.S.B., C.W.L., and A.E.K. acknowledge support from the John D. and Catherine T. MacArthur Foundation, the Alfred P. Sloan Foundation, the ISI Foundation, the Paul Allen Foundation, the Army Research Laboratory (W911NF-10-2-0022), the Army Research Office (Bassett-W911NF-14-1-0679, Grafton-W911NF-16-1-0474, DCIST- W911NF-17-2-0181), the Office of Naval Research, the National Institute of Mental Health (2-R01-DC-009209-11, R01-MH112847, R01-MH107235, R21-M MH-106799), the National Institute of Child Health and Human Development (1R01HD086888-01), National Institute of Neurological Disorders and Stroke (R01 NS099348), and the National Science Foundation (BCS-1441502, BCS-1430087, NSF PHY-1554488 and BCS-1631550). The content is solely the responsibility of the authors and does not necessarily represent the official views of any of the funding agencies.
 
\item[Author contributions] C.W.L., A.E.K., and D.S.B. conceived the project. C.W.L. designed the model and performed the analysis. C.W.L., A.E.K., and D.S.B. planned the experiments and discussed the results. C.W.L., A.E.K., and N.N. performed the experiments. C.W.L. wrote the manuscript and Supplementary Information. A.E.K., N.N., and D.S.B. edited the manuscript and Supplementary Information.
 
\item[Competing interests] The authors declare no competing interests.
 
\item[Corresponding author] Correspondence and requests for materials should be addressed to D.S.B. \\ (dsb@seas.upenn.edu).
 
\item[Supplementary information] Supplementary text and figures accompany this paper.
\end{addendum}

\newpage

\noindent {\Large \myfont \textbf{Supplementary Information}}

\noindent {\large \myfont \textbf{\textit{Abstract representations of events arise from mental errors in learning and memory}}}

\section{Introduction}

In this Supplementary Information, we provide extended discussion and data to support the results presented in the main text. The content is organized to roughly mirror the organization of the paper. In Sec. \ref{node_effects}, we present experimental evidence that human reaction times -- in addition to depending on higher-order network features -- also reflect differences in fine-scale structure at the level of individual nodes. Just as for the higher-order effects presented in the main text, we demonstrate that these fine-scale phenomena are accurately predicted by our maximum entropy model. In Sec. \ref{measuring_effects}, we present the mixed effects models that were used to estimate the cross-cluster surprisal and modular-lattice effects. In Sec. \ref{recency}, we demonstrate that the cross-cluster surprisal and modular-lattice effects cannot simply be explained by recency by directly controlling for the recency of stimuli. In Sec. \ref{Hamiltonian}, we use Hamiltonian walks to experimentally control for recency. In Sec. \ref{early_trials}, we show that the cross-cluster surprisal and modular-lattice effects persist even when considering all 1500 trials for each subject. In Sec. \ref{errors}, we show that the probability of an error on the serial response tasks increases for between- versus within-cluster transitions in the modular graph, indicating that the free energy framework can be used to predict human behaviors beyond reaction times. In Sec. \ref{violations}, we present the mixed effects models that were used to estimate the effects of violations in the ring graph. In Sec. \ref{violations_recency}, we show that the effects of network violations cannot be explained by recency alone. In Sec. \ref{forgetting}, we discuss why the forgetting of past stimuli altogether cannot explain the higher-order network effects that we examine in the main text. In Sec. \ref{gradient}, to aid in the reconstruction of our gradient descent algorithm for estimating the inverse temperature $\beta$ from subjects' reaction times, we derive an analytic form for the gradient of the RMS prediction error of our model with respect to $\beta$. In Sec. \ref{successor}, we discuss the relationship between our model and the successor representation in reinforcement learning.

\section{The effects of node heterogeneity on human expectations}
\label{node_effects}

In the main text, we demonstrated that human expectations depend critically on the higher-order network structure of transitions. In addition to these higher-order phenomena, it has long been known that human expectations also reflect differences in the fine-scale structure of transition networks.\cite{Fiser-01, Kahn-01} For instance, humans are surprised by rare transitions, represented in a transition network by edges with low probability weight.\cite{Saffran-01} Here, we provide empirical evidence showing that people's expectations also depend on the local topologies of the nodes that bookend a transition, and that these fine-scale effects are consistently predicted by our maximum entropy model.

\begin{figure}[t!]
\centering
\includegraphics[width = .8\textwidth]{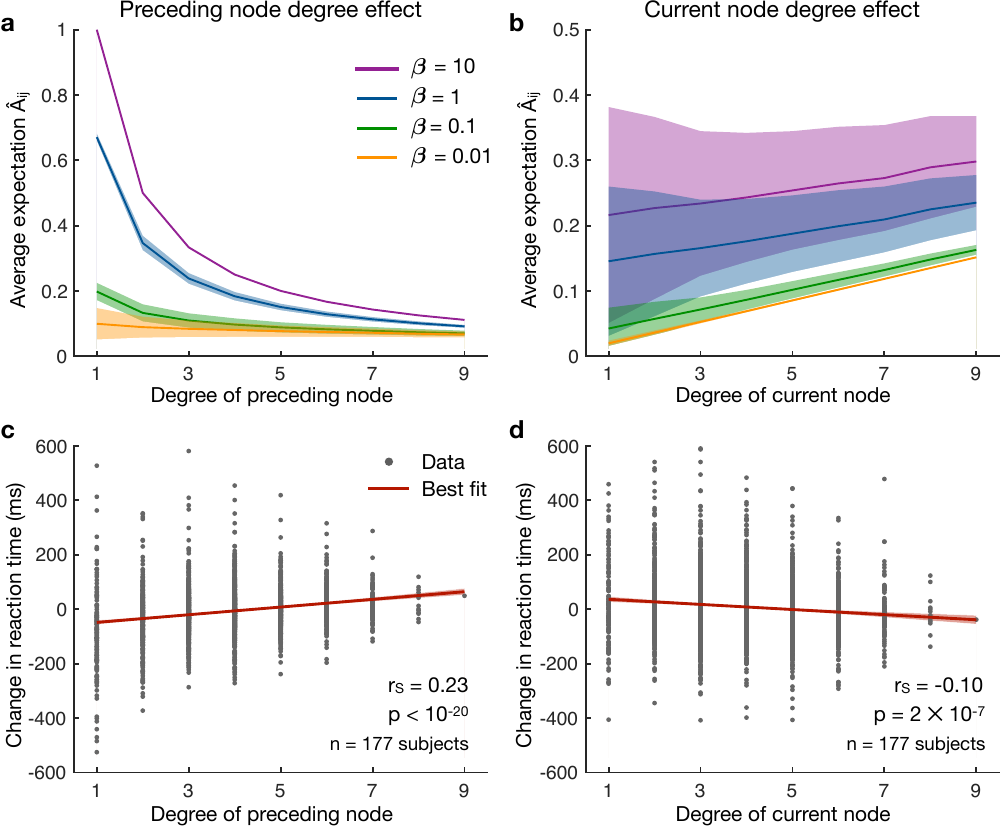} \\
\raggedright
\myfont\textbf{Supplementary Fig. 1: The effects of node degree on reaction times.}
\captionsetup{labelformat=empty}
{\spacing{1.25} \caption{\myfont \textbf{a}, The average expectation $\hat{A}_{ij}$ plotted with respect to the degree of the preceding node $i$ across a range of inverse temperatures $\beta$. As expected, expectations decrease as the degree of the preceding node increases; and for $\beta = 10$, we have $\hat{A}_{ij} \approx A_{ij} = 1/k_i$. The lines and shaded regions represent averages and 95\% confidence intervals over 1000 randomly-generated Erd\"{o}s-R\'{e}nyi networks. \textbf{b}, People exhibit sharp increases in reaction time following nodes of higher degree, with Spearman's correlation $r_S = 0.23$. The data is combined across 177 subjects, each of whom was asked to respond to a sequence of 1500 stimuli drawn from a random Erd\"{o}s-R\'{e}nyi network. Each data point represents the average reaction time for one node of a graph, and so each subject contributes 15 points. The line and shaded region represent the best fit and 95\% confidence interval, respectively. \textbf{c}, The average expectation $\hat{A}_{ij}$ plotted with respect to the degree of the current node $j$ across the same range of inverse temperatures as in \textbf{a}. \textbf{d}, People exhibit a steady decline in reaction times as the current node degree increases, with Spearman's correlation $r_S = -0.10$. Source data are provided as a Source Data file.}}
\end{figure}

In order to clearly study the effects of higher-order network structure, in the main text we focused on networks with uniform edge weights and node degrees. Here, to study the effects of node heterogeneity, we instead consider a set of Erd\"{o}s-R\'{e}nyi random graphs with the same number of nodes ($N=15$) and edges ($30$) as in our previous modular and lattice graphs. To ensure that the random walks are properly defined, we set the transition probability $A_{ij}$ of each edge in the graph to $1/k_i$, where $k_i$ is the degree of node $i$. Since the probabilities $A_{ij}$ decrease as the degree $k_i$ increases, one should suspect that high-degree (or hub) nodes yield decreased anticipations -- and therefore increased reaction times -- at the next step of a random sequence. Indeed, using Eq. (6) from the main text, we find that our model analytically predicts decreased expectations following a high-degree node (Supplementary Fig. 1a). Furthermore, across 177 human subjects, we find a strong positive correlation between people's reaction times and the degree of the preceding node in the sequence (Supplementary Fig. 1b).

Interestingly, while people's anticipations exhibit a sharp decline if the preceding node has high-degree, our model predicts that these hub nodes instead yield increased anticipations on the current step (Supplementary Fig. 1c). Thus, while hub nodes give rise to marked increases in reaction times on the subsequent step, these high-degree nodes actually yield faster reactions on the current step\cite{Kahn-01} (Supplementary Fig. 1d). This juxtaposition of effects from one time step to the next highlights the complex ways in which the network structure of transitions can affect people's mental representations. Additionally, the success of our model in predicting these competing phenomena further strengthens our conclusion that mental errors play a crucial role in shaping people's internal expectations.

\section{Measuring higher-order network effects}
\label{measuring_effects}

In order to extract the effects of higher-order network structure on subjects' reaction times, we use linear mixed effects models, which have become prominent in human research where many measurements are made for each subject.\cite{Schall-01,Baayen-01} To fit our mixed effects models and to estimate the statistical significance of each effect we use the \texttt{fitlme} function in MATLAB (R2018a). In what follows, when referring to our mixed effects models, we adopt the standard R notation.\cite{Bates-01}

\subsection{Cross-cluster surprisal effect}

We first measure the cross-cluster surprisal effect (Fig. 2a) using a mixed effects model with the formula `$\text{RT}\sim \log(\text{Trial})*\text{Stage} + \text{Target} + \text{Recency} + \text{Trans}\_\text{Type} + (1 + \log(\text{Trial})*\text{Stage} + \text{Recency} + \text{Trans}\_\text{Type} \,|\, \text{ID})$', where RT is the reaction time, Trial is the trial number between 501 and 1500, Stage is the stage of the experiment (either one or two), Target is the target button combination, Recency is the number of trials since last observing a node,\cite{Baddeley-01} Trans$\_$Type is the type of transition (either within-cluster or between-cluster), and ID is each subject's unique ID. We remark that our inclusion of Recency in the model is intended to distinguish the graph effects that we are interested in studying from the possible confound of recency, an effect that we directly control for in Sec. \ref{recency}. The mixed effects model is summarized in Supplementary Table 1, reporting a 35 ms increase in reaction times for between-cluster transitions relative to within-cluster transitions (Fig. 2a). This result is measured from the reaction time data for all 173 subjects that observed random walks in the modular graph.

\begin{figure}[t!]
\centering
\begin{tabular}{|c|c|c|c|c|}
\hline
Effect & Estimate (ms) & t-value & Pr$(>$$|$t$|)$ & Significance \\
\hline
\hline
(Intercept) & $1418.9 \pm 73.1$ & $19.42$ & $< 0.001$ & $***$ \\
\hline
$\log(\text{Trial})$ & $-92.1\pm 9.2$ & $-9.96$ & $< 0.001$ & $***$ \\
\hline
Stage & $-551.5 \pm 85.0$ & $-6.48$ & $< 0.001$ & $***$ \\
\hline
Recency & $1.4 \pm 0.1$ & $23.57$ & $< 0.001$ & $***$ \\
\hline
\rowcolor{LightGrey}
Trans$\_$Type & $34.9 \pm 6.0$ & $5.77$ & $< 0.001$ & $***$ \\
\hline
$\log(\text{Trial})$:Stage & $67.0 \pm 11.4$ & $5.89$ & $< 0.001$ & $***$ \\
\hline
\end{tabular}
\vskip 12pt
\raggedright
\captionsetup{labelformat=empty}
{\spacing{1.25} \caption{\myfont \textbf{Supplementary Table 1: Mixed effects model measuring the cross-cluster surprisal effect.} A mixed effects model fit to the reaction time data for the modular graph with the goal of measuring the cross-cluster surprisal effect. We find a significant 35 ms increase in reaction times (173 subjects) for between-cluster transitions versus within-cluster transitions (grey). The significance column represents $p$-values less than 0.001 ($***$), less than 0.01 ($**$), and less than 0.05 ($*$). Source data are provided as a Source Data file.}}
\end{figure}

\subsection{Modular-lattice effect}

We next measure the modular-lattice effect (Fig. 2b) using a mixed effects model of the form `$\text{RT}\sim \log(\text{Trial})*\text{Stage} + \text{Target} + \text{Recency} + \text{Graph} + (1 + \log(\text{Trial})*\text{Stage} + \text{Recency} \,|\, \text{ID})$', where Graph represents the type of transition network, either modular or lattice. Note that we only include Graph as a fixed effect because the corresponding mixed effect is not statistically significant. The mixed effects model is summarized in Supplementary Table 2, reporting a 23 ms increase in reaction times in the lattice graph relative to the modular graph (Fig. 2b). This result is measured from the reaction time data for the 72 subjects that observed random walks in both the modular and lattice graphs.

\begin{figure}[t!]
\centering
\begin{tabular}{|c|c|c|c|c|}
\hline
Effect & Estimate (ms) & t-value & Pr$(>$$|$t$|)$ & Significance \\
\hline
\hline
(Intercept) & $1436.5 \pm 48.4$ & $29.67$ & $< 0.001$ & $***$ \\
\hline
$\log(\text{Trial})$ & $-97.2 \pm 6.1$ & $-15.89$ & $< 0.001$ & $***$ \\
\hline
Stage & $-555.2 \pm 59.3$ & $-9.36$ & $< 0.001$ & $***$ \\
\hline
Recency & $1.7 \pm 0.1$ & $29.94$ & $< 0.001$ & $***$ \\
\hline
\rowcolor{LightGrey}
Graph & $22.8 \pm 5.8$ & $3.95$ & $< 0.001$ & $***$ \\
\hline
$\log(\text{Trial})$:Stage & $71.4 \pm 8.4$ & $8.48$ & $< 0.001$ & $***$ \\
\hline
\end{tabular}
\vskip 12pt
\raggedright
\captionsetup{labelformat=empty}
{\spacing{1.25} \caption{\myfont \textbf{Supplementary Table 2: Mixed effects model measuring the modular-lattice effect.} A mixed effects model fit to the reaction time data for the modular and lattice graphs with the goal of measuring the modular-lattice effect. We find a significant 23 ms increase in reaction times overall (72 subjects) in the lattice graph relative to the modular graph (grey). The significance column represents $p$-values less than 0.001 ($***$), less than 0.01 ($**$), and less than 0.05 ($*$). Source data are provided as a Source Data file.}}
\end{figure}

\section{Cross-cluster surprisal with Hamiltonian walks}
\label{Hamiltonian}

\begin{figure}[t!]
\centering
\begin{tabular}{|c|c|c|c|c|}
\hline
Effect & Estimate (ms) & t-value & Pr$(>$$|$t$|)$ & Significance \\
\hline
\hline
(Intercept) & $1420.8 \pm 162.3$ & $8.75$ & $< 0.001$ & $***$ \\
\hline
$\log(\text{Trial})$ & $-101.4 \pm 22.7$ & $-4.48$ & $< 0.001$ & $***$ \\
\hline
Recency & $0.6 \pm 0.1$ & $5.00$ & $< 0.001$ & $***$ \\
\hline
\rowcolor{LightGrey}
Trans$\_$Type & $35.6 \pm 13.7$ & $2.59$ & $0.010$ & $**$ \\
\hline
\end{tabular}
\vskip 12pt
\raggedright
\captionsetup{labelformat=empty}
{\spacing{1.25} \caption{\myfont \textbf{Supplementary Table 3: Mixed effects model measuring the cross-cluster surprisal effect in Hamiltonian walks.} A mixed effects model fit to subjects' reaction times in Hamiltonian walks on the modular graph with the goal of measuring the cross-cluster surprisal effect. We find a significant 36 ms increase in reaction times (120 subjects) for between-cluster transitions versus within-cluster transitions (grey). The significance column represents $p$-values less than 0.001 ($***$), less than 0.01 ($**$), and less than 0.05 ($*$). Source data are provided as a Source Data file.}}
\end{figure}

Throughout the main text, we assume that people's reaction times reflect their internal representations of the transition structure. To justify this assumption, we must show that the higher-order network effects cannot simply be explained by recency. Here, we measure the cross-cluster surprisal effect while experimentally controlling for recency using Hamiltonian walks. In contrast to random walks, Hamiltonian walks visit each node in the transition graph exactly once, thereby guaranteeing that each node is visited once every 15 trials. We run a new experiment in which each subject (out of 120 subjects) is presented with a sequence of 1500 stimuli drawn from the modular graph: The first 700 nodes reflect a standard random walk, while the remaining 800 trials consist of 8 repeated segments of 85 stimuli specified by a random walk followed by 15 stimuli specified by a Hamiltonian walk. The initial 700 random walk trials are meant to constitute a learning phase in which the subject builds an internal representation of the modular graph. Since, in the modular graph, Hamiltonian walks do not obey the same transition probabilities as random walks, the sequences of 85 random walk trials between each Hamiltonian sequence are meant to help the subject maintain their learned representation. Within the set of Hamiltonian walks through the modular graph, the probability of transitioning from one cluster boundary node to the adjacent one (if not already visited) is 1, whereas the probability of transitioning from the latter boundary node to each of the adjacent non-boundary nodes is 1/3. To eliminate this difference, we randomly selected one fixed Hamiltonian walk for each subject. This fixed walk was entered at a different node depending on where the preceding walk terminated, and we randomly switched between forward and backward traversals for each walk.\cite{Schapiro-01}

We measure the cross-cluster surprisal within the Hamiltonian trials using a mixed effects model with the formula `$\text{RT} \sim \log(\text{Trial}) + \text{Target} + \text{Recency} + \text{Trans}\_\text{Type} + (1 + \log(\text{Trial}) + \text{Recency} + \text{Trans}\_\text{Type} \,|\,\text{ID})$', where each of the variables has been defined previously. The model is summarized in Supplementary Table 3, reporting a 36 ms increase in reaction times for between-cluster transitions relative to within-cluster transitions within Hamiltonian trials ($p = 0.010$), matching (within errors) the effect size reported in the original experiment that only included random walks (see Supplementary Table 1). This result is measured from the reaction time data for all 120 subjects that observed random walks with Hamiltonian walks interspersed in the modular graph. This result indicates that the cross-cluster surprisal effect cannot be explained by recency alone, and must therefore must be at least partially driven by people's internal representations of the transition structure.

\subsection{Removing Hamiltonian trials before the first cross-cluster transition}

The purpose of the Hamiltonian walk experiment described above is to experimentally control for the effects of recency on people's reaction times. However, when thinking carefully about the transition from a random walk to a Hamiltonian walk, it becomes clear that recency might still have a noticeable impact. Consider, for example the last few trials of a random walk preceding a transition to a Hamiltonian walk -- the corresponding stimuli are likely to belong to the same module in the modular graph. When the sequence converts to a Hamiltonian walk, the first few stimuli are also likely to belong to the same module, thereby inducing a decrease in reaction times due to recency. Therefore, in order to more thoroughly control for recency effects, we considered only trials after the first cross-cluster transition within each Hamiltonian walk. We carry out this restricted analysis using the same form for the mixed effects model as that described above: `$\text{RT} \sim \log(\text{Trial}) + \text{Target} + \text{Recency} + \text{Trans}\_\text{Type} + (1 + \log(\text{Trial}) + \text{Recency} + \text{Trans}\_\text{Type} \,|\,\text{ID})$'. The model, which is summarized in Supplementary Table 4, estimates a significant cross-cluster surprisal effect of $28$ ms ($p = 0.043$), again matching within errors the effect size found in the original random walk data.

\begin{figure}[t!]
\centering
\begin{tabular}{|c|c|c|c|c|}
\hline
Effect & Estimate (ms) & t-value & Pr$(>$$|$t$|)$ & Significance \\
\hline
\hline
(Intercept) & $1536.6 \pm 178.0$ & $8.63$ & $< 0.001$ & $***$ \\
\hline
$\log(\text{Trial})$ & $-116.3 \pm 24.8$ & $-4.68$ & $< 0.001$ & $***$ \\
\hline
Recency & $0.4 \pm 0.1$ & $3.00$ & $0.003$ & $**$ \\
\hline
\rowcolor{LightGrey}
Trans$\_$Type & $28.2 \pm 13.9$ & $2.03$ & $0.043$ & $*$ \\
\hline
\end{tabular}
\vskip 12pt
\raggedright
\captionsetup{labelformat=empty}
{\spacing{1.25} \caption{\myfont \textbf{Supplementary Table 4: Mixed effects model measuring the cross-cluster surprisal effect in restricted Hamiltonian walks.} A mixed effects model fit to subjects' reaction times after the first cross-cluster transition within each Hamiltonian walk. We find a significant 28 ms increase in reaction times (120 subjects) for between-cluster transitions versus within-cluster transitions (grey). The significance column represents $p$-values less than 0.001 ($***$), less than 0.01 ($**$), and less than 0.05 ($*$). Source data are provided as a Source Data file.}}
\end{figure}

\subsection{Decreasing cross-cluster surprisal with increasing Hamiltonian trials}

As discussed above, the first 700 trials of each sequence were drawn from a random walk to allow subjects to build an internal representation of the random walk transition structure. Since the transition probabilities reflected in the Hamiltonian walks differ from those in the random walks, we expect subjects' representations of the transition structure to shift as they observe increasing numbers of Hamiltonian trials. Therefore, to further establish the notion that people's reactions are primarily driven by their internal representations, here we show that the strength of the cross-cluster surprisal decreases as subjects observe increasing numbers of Hamiltonian trials. To do so, we use a mixed effects model with the formula `$\text{RT} \sim \log(\text{Trial})*\text{Trans}\_\text{Type} + \text{Target} + \text{Recency} + (1 + \log(\text{Trial}) + \text{Recency} + \text{Trans}\_\text{Type} \,|\,\text{ID})$', where the only difference with the formula above is that here we include an interaction term between $\log(\text{Trial})$ and Trans$\_$Type. The results of the fitted model are summarized in Supplementary Table 5, reporting a significant decrease in the strength of the cross-cluster surprisal with increasing Hamiltonian trials ($p = 0.024$).

\begin{figure}[t!]
\centering
\begin{tabular}{|c|c|c|c|c|}
\hline
Effect & Estimate (ms) & t-value & Pr$(>$$|$t$|)$ & Significance \\
\hline
\hline
(Intercept) & $1394.1\pm 188.8$ & $7.39$ & $< 0.001$ & $***$ \\
\hline
$\log(\text{Trial})$ & $-96.1 \pm 26.4$ & $-3.64$ & $< 0.001$ & $***$ \\
\hline
Recency & $0.4 \pm 0.1$ & $3.02$ & $0.003$ & $**$ \\
\hline
Trans$\_$Type & $640.3 \pm 271.9$ & $2.35$ & $0.019$ & $*$ \\
\hline
\rowcolor{LightGrey}
$\log(\text{Trial})$:Trans$\_$Type & $-87.2 \pm 38.7$ & $-2.25$ & $0.024$ & $*$ \\
\hline
\end{tabular}
\vskip 12pt
\raggedright
\captionsetup{labelformat=empty}
{\spacing{1.25} \caption{\myfont \textbf{Supplementary Table 5: Mixed effects model measuring the decrease in cross-cluster surprisal with increasing Hamiltonian trials.} A mixed effects model fit to subjects' reaction times in Hamiltonian walks on the modular graph with the goal of measuring the dependence of the cross-cluster surprisal on increasing trial number. We find a significant decrease in the strength of the cross-cluster surprisal with increasing trials (grey), indicating that the introduction of Hamiltonian walks weakens people's internal representations of the random walk structure (120 subjects). The significance column represents $p$-values less than 0.001 ($***$), less than 0.01 ($**$), and less than 0.05 ($*$). Source data are provided as a Source Data file.}}
\end{figure}

\subsection{Experimental setup and procedures}

Subjects performed a self-paced serial reaction time task, as described in the Methods section of the main text. The only difference between this experiment and the original random walk experiments is that the 1500 trials were split into 700 trials drawn as a random walk and a subsequent 800 trials divided into 8 segments of 85 random walk trials followed by 15 Hamiltonian walk trials, all drawn from the modular graph. In total, we recruited 120 subjects to perform this Hamiltonian walk experiment, and they were paid up to \$5 each for an estimated 30 minutes: \$3.50 for completing the task and \$1.50 for correctly responding on at least 90\% of the trials.

\section{Controlling for recency in random walks}
\label{recency}

In the previous section, we showed that cross-cluster surprisal remains significant during Hamiltonian walks, which experimentally control for the recency of stimuli. Building upon this result, in this section we measure the cross-cluster surprisal and modular-lattice effects in our initial random walk data while filtering our data based on stimulus recency.

\subsection{Cross-cluster surprisal effect while controlling for recency}

In order to control for recency, we filter our data to only include trials in which the current stimulus was last seen a specific number of trials in the past. For example, when studying trials with a recency of four, we only consider reaction times from our experiments for which the current stimulus was last seen four trials previously. After filtering the data, we then estimate the cross-cluster surprisal effect using a mixed effects model of the form `$\text{RT}\sim \log(\text{Trial})*\text{Stage} + \text{Target} + \text{Trans}\_\text{Type} + (1 + \log(\text{Trial})*\text{Stage} + \text{Trans}\_\text{Type} \,|\, \text{ID})$'. Supplementary Fig. 2a shows the estimated increase in reaction times for within-cluster versus between-cluster transitions after controlling for recency. Specifically, we consider recency values of two (the minimum) through nine, and we also consider trials with recency greater than or equal to 10, for which the effects of recency should be small. We remark that we do not include trials of recency three in our analysis because, due to the topology of the modular graph, there do not exist between-cluster transitions with recency three. We find significant effects for all recency values besides eight. 

\begin{figure}[t!]
\centering
\includegraphics[width = .7\textwidth]{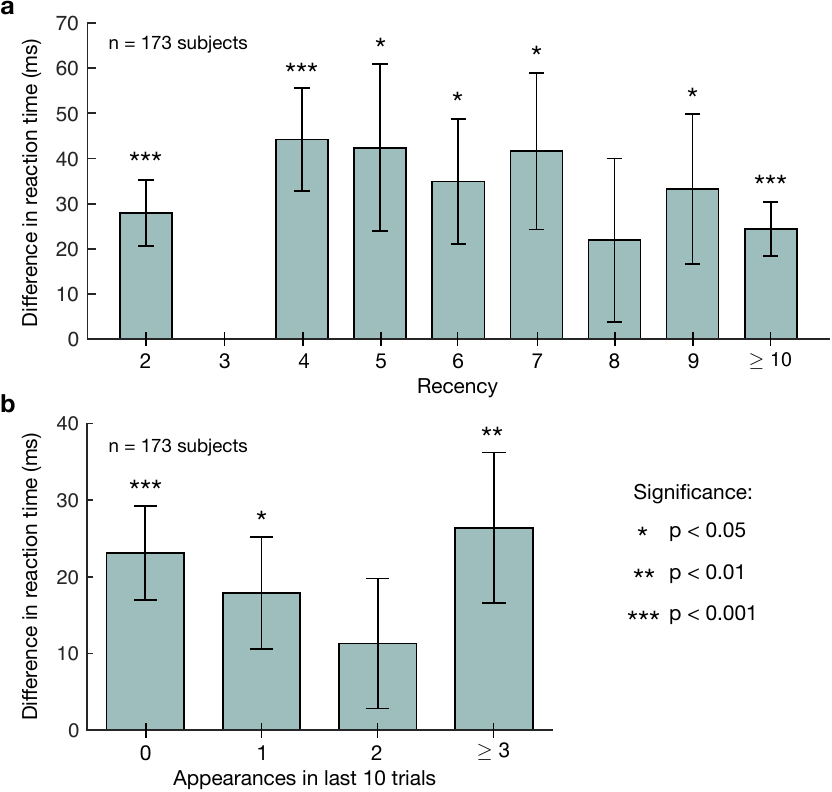} \\
\raggedright
\myfont\textbf{Supplementary Fig. 2: Cross-cluster surprisal while controlling for recency.}
\captionsetup{labelformat=empty}
{\spacing{1.25} \caption{\myfont \textbf{a}, Increase in reaction times for between-cluster versus within-cluster transitions in the modular graph after controlling for the recency of stimuli. We note that, due to the topology of the modular graph, there do not exist between-cluster transitions with recency three. We find significant cross-cluster surprisal effects for all recency values besides eight. \textbf{b}, Increase in reaction times for between- versus within-cluster transitions after controlling for the number of times that the current stimulus has appeared in the previous 10 trials. We observe significant cross-cluster surprisal for all numbers of recent stimulus appearances besides two. Effect sizes (represented by bar plots), standard deviations (represented by error bars), and \textit{p}-values are estimated using mixed effects models. The results are measured for all 173 subjects that observed random walks in the modular graph. Source data are provided as a Source Data file.}}
\end{figure}

In addition to controlling for the recency of stimuli, we also study the cross-cluster surprisal while controlling for the number of appearances of the current stimulus in the last 10 trials. In particular, we filter our data to only include trials for which the current stimulus was seen a specified number of times in the previous 10 trials, and for each set of filtered data we estimate the cross-cluster surprisal using a mixed effects model of the same form as above. We observe a significant increase in reaction times for between- versus within-cluster transitions for all trials except for those for which the stimulus appeared twice in the last 10 trials (Supplementary Fig. 2b). Together, these results demonstrate that the cross-cluster surprisal effect cannot be explained by recency alone, and therefore must stem, at least in part, from people's internal representations of the transition structure.

\subsection{Modular-lattice effect while controlling for recency}

We next consider the modular-lattice effect after controlling for recency. Filtering the data from the modular and lattice graphs to only include trials of a given recency, we estimate the difference in reaction times between the two graphs using a mixed effects model of the form `$\text{RT}\sim \log(\text{Trial})*\text{Stage} + \text{Target} + \text{Graph} + (1 + \log(\text{Trial})*\text{Stage} \,|\, \text{ID})$'. Supplementary Fig. 3a shows that we find a significant increase in reaction times for the lattice graph relative to the modular graph for all recency values considered besides three, nine, and $\ge$ 10. Additionally, in Supplementary Fig. 3b, we control for the number of appearances of the current stimulus in the previous 10 trials. Using a mixed effects model of the same form as that above, we find a significant modular-lattice effect in two of the four conditions. Together, these results demonstrate that the difference in reaction times between the modular and lattice graphs persists after controlling for the recency of stimuli, indicating that people are better able to anticipate transitions in the modular graph than in the lattice graph.

\begin{figure}[t!]
\centering
\includegraphics[width = .7\textwidth]{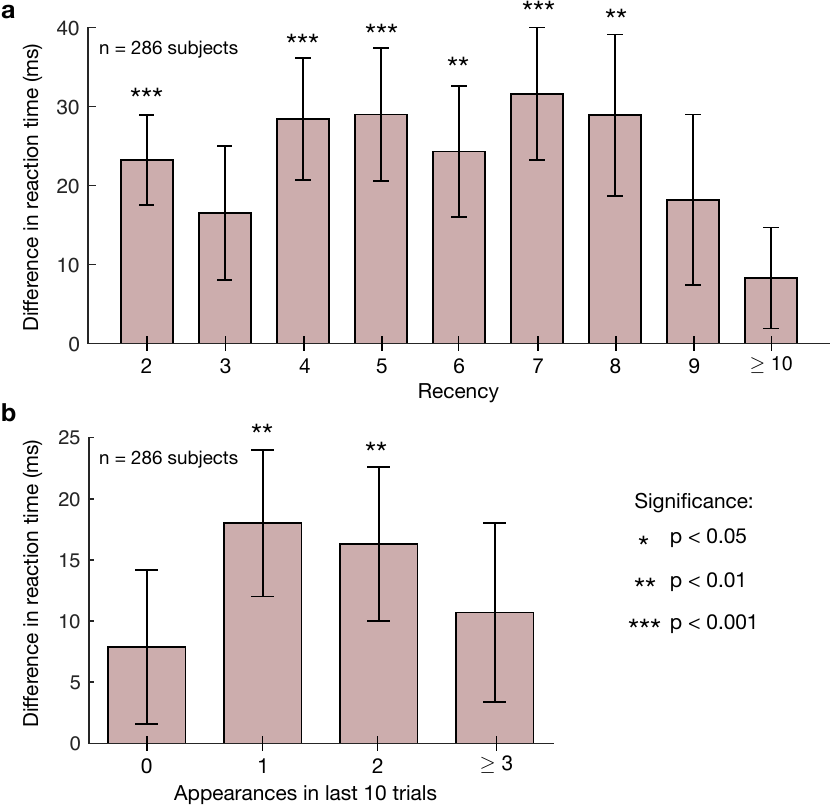} \\
\raggedright
\myfont\textbf{Supplementary Fig. 3: Modular-lattice effect while controlling for recency.}
\captionsetup{labelformat=empty}
{\spacing{1.25} \caption{\myfont \textbf{a}, Difference in reaction times between the lattice and modular graphs after controlling for the recency of stimuli. We observe a significant increase in reaction times for the lattice graph relative to the modular graph for all recency values besides three, nine, and $\ge$ 10. \textbf{b}, Difference in reaction times between the lattice and modular graphs after controlling for the number of times the current stimulus has appeared in the previous 10 trials. We find a significant modular-lattice effect for one and two stimulus appearances in the last 10 trials. Effect sizes (represented by bar plots), standard deviations (represented by error bars), and \textit{p}-values are estimated using mixed effects models. The results are measured for all 72 subjects that observed random walks in both the modular and lattice graphs. Source data are provided as a Source Data file.}}
\end{figure}

\section{Measuring network effects including early trials}
\label{early_trials}

Throughout the above analysis of the serial response tasks, we purposefully omitted the first 500 trials for each subject, choosing instead to focus on the final 1000 trials. We did this in order to allow the subjects to build an internal representation of each network structure before probing their anticipations of transitions. Here, we show that this data processing step is not necessary to observe higher-order network effects; that is, we show that there exist significant network effects even if we include the first 500 trials in our analysis.

\subsection{Cross-cluster surprisal effect with early trials}

We first consider the cross-cluster surprisal effect defined by an increase in reaction times for transitions between clusters relative to transitions within clusters in the modular graph. Using a mixed effects model of the same form as that used in the previous analysis in Sec. \ref{measuring_effects} (i.e., `$\text{RT}\sim \log(\text{Trial})*\text{Stage} + \text{Target} + \text{Recency} + \text{Trans}\_\text{Type} + (1 + \log(\text{Trial})*\text{Stage} + \text{Recency} + \text{Trans}\_\text{Type} \,|\, \text{ID})$'), and including all 1500 trials for each subject, we find a significant 35 ms increase in reaction times for between- versus within-cluster transitions (Supplementary Table 6). We note that this effect is even larger than that observed in our previous analysis (Supplementary Table 1).

\begin{figure}[t!]
\centering
\begin{tabular}{|c|c|c|c|c|}
\hline
Effect & Estimate (ms) & t-value & Pr$(>$$|$t$|)$ & Significance \\
\hline
\hline
(Intercept) & $1340.0 \pm 44.4$ & $30.19$ & $< 0.001$ & $***$ \\
\hline
$\log(\text{Trial})$ & $-88.7 \pm 5.0$ & $-17.04$ & $< 0.001$ & $***$ \\
\hline
Stage & $-473.1 \pm 47.6$ & $-9.93$ & $< 0.001$ & $***$ \\
\hline
Recency & $1.5 \pm 0.1$ & $24.65$ & $< 0.001$ & $***$ \\
\hline
\rowcolor{LightGrey}
Trans$\_$Type & $35.4 \pm 6.0$ & $5.94$ & $< 0.001$ & $***$ \\
\hline
$\log(\text{Trial})$:Stage & $60.4 \pm 5.5$ & $11.06$ & $< 0.001$ & $***$ \\
\hline
\end{tabular}
\vskip 12pt
\raggedright
\captionsetup{labelformat=empty}
{\spacing{1.25} \caption{\myfont \textbf{Supplementary Table 6: Mixed effects model measuring the cross-cluster surprisal effect including the first 500 trials.} A mixed effects model fit to all of the reaction time data, including the first 500 trials for each subject, for the modular graph with the goal of measuring the cross-cluster surprisal effect. We find a significant 35 ms increase in reaction times (173 subjects) for between-cluster transitions versus within-cluster transitions. The significance column represents $p$-values less than 0.001 ($***$), less than 0.01 ($**$), and less than 0.05 ($*$). Source data are provided as a Source Data file.}}
\end{figure}

\subsection{Modular-lattice effect with early trials}

We next consider the modular-lattice effect defined by an increase in reaction times in the lattice graph relative to the modular graph. Using a mixed effects model of the same form as that used in the previous analysis in Sec. \ref{measuring_effects} (i.e., `$\text{RT}\sim \log(\text{Trial})*\text{Stage} + \text{Target} + \text{Recency} + \text{Graph} + (1 + \log(\text{Trial})*\text{Stage} + \text{Recency} \,|\, \text{ID})$'), and including all 1500 trials for each subject, we find a significant 16 ms increase in reaction times in the lattice versus the modular graph (Supplementary Table 7). These results demonstrate that higher-order network effects studied in the main text exist throughout the entire serial response task.

\begin{figure}[t!]
\centering
\begin{tabular}{|c|c|c|c|c|}
\hline
Effect & Estimate (ms) & t-value & Pr$(>$$|$t$|)$ & Significance \\
\hline
\hline
(Intercept) & $1357.0 \pm 30.3$ & $44.79$ & $< 0.001$ & $***$ \\
\hline
$\log(\text{Trial})$ & $-87.8 \pm 3.4$ & $-26.06$ & $< 0.001$ & $***$ \\
\hline
Stage & $-490.7 \pm 25.3$ & $-19.38$ & $< 0.001$ & $***$ \\
\hline
Recency & $2.0 \pm 0.1$ & $32.35$ & $< 0.001$ & $***$ \\
\hline
\rowcolor{LightGrey}
Graph & $16.3 \pm 5.4$ & $3.00$ & $0.003$ & $**$ \\
\hline
$\log(\text{Trial})$:Stage & $62.7 \pm 3.5$ & $17.76$ & $< 0.001$ & $***$ \\
\hline
\end{tabular}
\vskip 12pt
\raggedright
\captionsetup{labelformat=empty}
{\spacing{1.25} \caption{\myfont \textbf{Supplementary Table 7: Mixed effects model measuring the modular-lattice effect including the first 500 trials.} A mixed effects model fit to all of the reaction time data, including the first 500 trials for each subject, for the modular and lattice graphs with the goal of measuring the modular-lattice effect. We find a significant 16 ms increase in reaction times overall (72 subjects) in the lattice graph relative to the modular graph. The significance column represents $p$-values less than 0.001 ($***$), less than 0.01 ($**$), and less than 0.05 ($*$). Source data are provided as a Source Data file.}}
\end{figure}

\section{Network effects on error trials}
\label{errors}

Thus far we have focused on predicting human reaction times as a proxy for people's anticipations of transitions. Another way to probe anticipation is by studying the trials on which subjects respond incorrectly; one might expect that the probability of an erroneous response should increase with decreasing anticipation. Here, we test this hypothesis for between- versus within-cluster transitions in the modular graph and for all transitions in the modular graph versus the lattice graph.

\subsection{Cross-cluster surprisal effect on errors}

First, we consider the cross-cluster surprisal effect on errors defined by an increase in task errors for transitions between clusters relative to transitions within clusters in the modular graph. We employ a mixed effects model with formula `$\text{Error} \sim \log(\text{Trial}) + \text{Stage} + \text{Target} + \text{Recency} + \text{Trans}\_\text{Type} + (1 + \log(\text{Trial})\,|\,\text{ID})$', where Error indicates whether the subject provided an incorrect (`1') or correct (`0') response. Note that, relative to our measurement of the cross-cluster surprisal for reaction times in Sec. \ref{measuring_effects}, we have removed the fixed effect interaction between $\log(\text{Trial})$ and Stage as well as the mixed effects for the variables Stage, Recency, and Trans$\_$Type because they are not statistically significant in this setting. We find a significant increase in errors for between- versus within-cluster transitions (Supplementary Table 8), suggesting yet again that subjects have weaker anticipation for cross-cluster transitions than for within-cluster transitions.

\begin{figure}[t!]
\centering
\begin{tabular}{|c|c|c|c|c|}
\hline
Effect & Estimate & t-value & Pr$(>$$|$t$|)$ & Significance \\
\hline
\hline
(Intercept) & $0.005 \pm 0.012$ & $0.39$ & $0.697$ &  \\
\hline
$\log(\text{Trial})$ & $0.004 \pm 0.002$ & $2.14$ & $0.032$ & $*$ \\
\hline
Stage & $0.015\pm 0.007$ & $2.14$ & $0.032$ & $*$  \\
\hline
Recency & $< 0.001$ & $16.52$ & $< 0.001$ & $***$  \\
\hline
\rowcolor{LightGrey}
Trans$\_$Type & $0.004 \pm 0.002$ & $2.83$ & $0.005$ & $**$ \\
\hline
\end{tabular}
\vskip 12pt
\raggedright
\captionsetup{labelformat=empty}
{\spacing{1.25} \caption{\myfont \textbf{Supplementary Table 8: Mixed effects model measuring the cross-cluster effect on task errors.} A mixed effects model fit to predict error trials for the modular graph with the goal of measuring the cross-cluster effect on task errors. We find a significant increase in task errors (173 subjects) for between-cluster transitions relative to within-cluster transitions (grey). The significance column represents $p$-values less than 0.001 ($***$), less than 0.01 ($**$), and less than 0.05 ($*$). Source data are provided as a Source Data file.}}
\end{figure}

\subsection{Modular-lattice effect on errors}

\begin{figure}[t!]
\centering
\begin{tabular}{|c|c|c|c|c|}
\hline
Effect & Estimate & t-value & Pr$(>$$|$t$|)$ & Significance \\
\hline
\hline
(Intercept) & $0.026 \pm 0.009$ & $3.05$ & $0.002$ & $**$ \\
\hline
$\log(\text{Trial})$ & $0.002 \pm 0.001$ & $1.47$ & $0.142$ &  \\
\hline
Stage & $0.003 \pm 0.003$ & $0.98$ & $0.325$ &  \\
\hline
Recency & $< 0.001$ & $14.62$ & $< 0.001$ & $***$ \\
\hline
\rowcolor{LightGrey}
Graph & $-0.004 \pm 0.003$ & $-1.34$ & $0.180$ &  \\
\hline
\end{tabular}
\vskip 12pt
\raggedright
\captionsetup{labelformat=empty}
{\spacing{1.25} \caption{\label{MLerrors} \myfont \textbf{Supplementary Table 9: Mixed effects model measuring the modular-lattice effect on task errors.} A mixed effects model fit to predict error trials for the modular and lattice graphs with the goal of measuring the modular-lattice effect on task errors. We do not find a significant change in errors based on the graph (grey; 72 subjects). The significance column represents $p$-values less than 0.001 ($***$), less than 0.01 ($**$), and less than 0.05 ($*$). Source data are provided as a Source Data file.}}
\end{figure}

Second, we consider the modular-lattice effect on errors defined by an increase in task errors for the lattice graph relative to the modular graph. We employ a mixed effects model with formula `$\text{Error} \sim \log(\text{Trial}) + \text{Stage} + \text{Target} + \text{Recency} + \text{Graph} + (1 + \log(\text{Trial}) + \text{Recency} + \text{Graph}\,|\,\text{ID})$', where each of the variables has been defined previously. We again note that we have removed the interaction between $\log(\text{Trial})$ and Stage because it was not statistically significant in our prediction of task errors. Inspecting the mixed effects model described in Supplementary Table 9, we do not find a significant difference in the number of task errors between the modular and lattice graphs. One possible explanation for this lack of an effect is that people's task accuracy is predominantly impacted by very poorly anticipated transitions. Thus, while anticipation in the lattice graph is lower than that in the modular graph on average, it could be the case that the significant decrease in anticipation for cross-cluster transitions in the modular graph yields a similar number of task errors overall.

\section{Measuring the effects of network violations}
\label{violations}

\begin{figure}[t!]
\centering
\begin{tabular}{|c|c|c|c|c|}
\hline
Effect & Estimate (ms) & t-value & Pr$(>$$|$t$|)$ & Significance \\
\hline
\hline
(Intercept) & $1352.7 \pm 79.2$ & $17.07$ & $< 0.001$ & $***$ \\
\hline
$\log(\text{Trial})$ & $-101.1 \pm 10.2$ & $-9.96$ & $< 0.001$ & $***$ \\
\hline
Recency & $2.1 \pm 0.1$ & $16.20$ & $< 0.001$ & $***$ \\
\hline
\rowcolor{LightGrey}
Top$\_$Dist (short vs. no violation) & $37.9 \pm 8.4$ & $4.50$ & $< 0.001$ & $***$ \\
\hline
\rowcolor{LightGrey}
Top$\_$Dist (long vs. no violation) & $63.3 \pm 7.8$ & $8.07$ & $< 0.001$ & $***$ \\
\hline
\end{tabular}
\vskip 12pt
\raggedright
\captionsetup{labelformat=empty}
{\spacing{1.25} \caption{\myfont \textbf{Supplementary Table 10: Mixed effects model measuring the effects of violations relative to standard transitions.} A mixed effects model fit to the reaction time data for the ring graph with the goal of measuring the effects of violations relative to standard transitions. We find a significant increase in reaction times of 38 ms (78 subjects) for short violations and 63 ms for long violations (grey), even after accounting for recency effects. The significance column represents $p$-values less than 0.001 ($***$), less than 0.01 ($**$), and less than 0.05 ($*$). Source data are provided as a Source Data file.}}
\end{figure}

We study the effects of violations of varying topological distance in the ring graph using a mixed effects model with the formula `$\text{RT}\sim \log(\text{Trial}) + \text{Target} + \text{Recency} + \text{Top}\_\text{Dist} + (1 + \log(\text{Trial}) + \text{Recency} + \text{Top}\_\text{Dist} \,|\, \text{ID})$', where Top$\_$Dist represents the topological distance of a transition, either one for a standard transition, two for a short violation, or three for a long violation. The results of fitting this mixed effects model are summarized in Supplementary Table 10, reporting increases in reaction times over standard transitions of 38 ms for short violations and 63 ms for long violations. Second, to measure the difference in reaction times between long and short violations, we implemented a model of the same form, but restricted Top$\_$Dist to only include short violations of topological distance two and long violations of topological distances three and four. This model is summarized in  Supplementary Table 11, reporting a 28 ms increase in reaction times for long violations relative to short violations. This result is measured from all 78 subjects that observed random walks with violations in the ring graph.

\begin{figure}[t!]
\centering
\begin{tabular}{|c|c|c|c|c|}
\hline
Effect & Estimate (ms) & t-value & Pr$(>$$|$t$|)$ & Significance \\
\hline
\hline
(Intercept) & $1380.9 \pm 156.1$ & $8.84$ & $< 0.001$ & $***$ \\
\hline
$\log(\text{Trial})$ & $-97.1 \pm 21.3$ & $-4.57$ & $< 0.001$ & $***$ \\
\hline
Recency & $0.7 \pm 0.3$ & $2.67$ & $0.008$ & $**$ \\
\hline
\rowcolor{LightGrey}
Top$\_$Dist (long vs. short violation) & $28.4 \pm 11.2 $ & $2.54$ & $0.011$ & $*$ \\
\hline
\end{tabular}
\vskip 12pt
\raggedright
\captionsetup{labelformat=empty}
{\spacing{1.25} \caption{\myfont \textbf{Supplementary Table 11: Mixed effects model measuring the effects of long versus short violations.} A mixed effects model fit to the reaction time data for the ring graph with the goal of measuring the effects of long versus short violations. We find a significant 28 ms increase in reaction times (78 subjects) for long violations relative to short violations (grey), even after accounting for recency effects. The significance column represents $p$-values less than 0.001 ($***$), less than 0.01 ($**$), and less than 0.05 ($*$). Source data are provided as a Source Data file.}}
\end{figure}

\begin{figure}
\centering
\includegraphics[width = .7\textwidth]{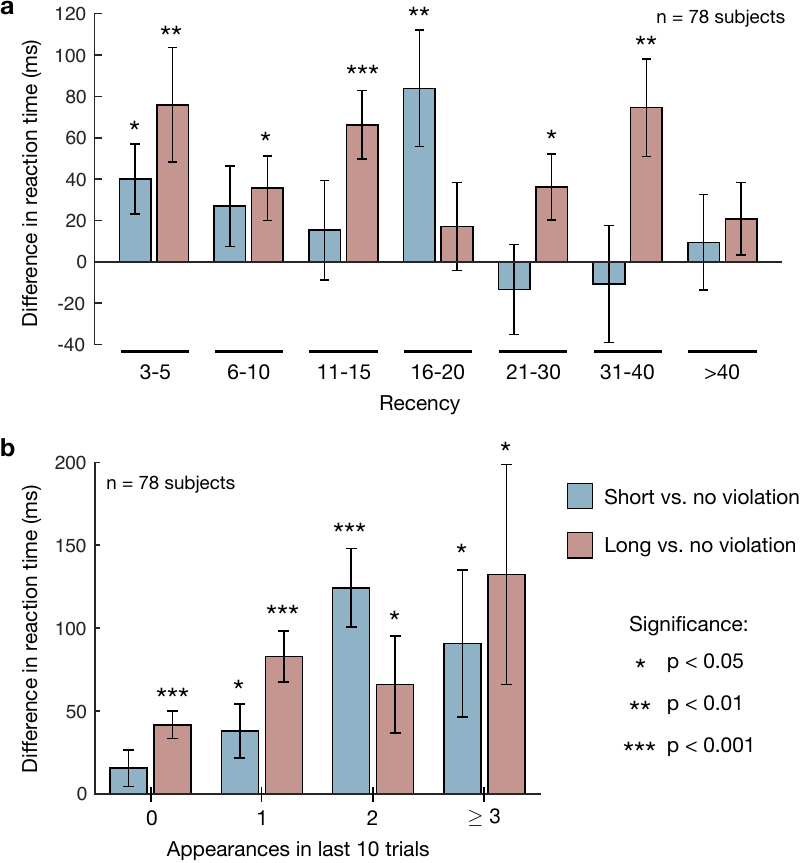} \\
\raggedright
\myfont\textbf{Supplementary Fig. 4: Comparing standard transitions to network violations while controlling for recency.}
\captionsetup{labelformat=empty}
{\spacing{1.25} \caption{\myfont \textbf{a}, Difference in reaction times between standard transitions and short violations (blue) or long violations (red) in the ring graph after controlling for the recency of stimuli. We observe at least one significant effect of network violations for all recency ranges less than 40. \textbf{b}, Increase in reaction times for short (blue) and long (red) network violations after controlling for the number of times the current stimulus has appeared in the previous 10 trials. For long violations, we find a significant increase in reaction times across all numbers of recent stimulus appearances. For short violations, we find a significant increase in reaction times across all numbers of recent stimulus appearances besides zero. Effect sizes (represented by bar plots), standard deviations (represented by error bars), and \textit{p}-values are estimated using mixed effects models. The results are measured for all 78 subjects that observed random walks with violations in the ring graph. Source data are provided as a Source Data file.}}
\end{figure}

\section{Controlling for recency: Network violations}
\label{violations_recency}

In the main text, we attribute the observed increase in reaction times for network violations to subjects' internal representations of the transition structure. Alternatively, these effects could be due to the fact that standard transitions are more likely than network violations to yield a stimulus that has been seen recently. To show that the effects of network violations are not simply driven by recency, we directly control for the recency of stimuli in our data. Because the violations data is more sparse than the standard random walk data (we only observe 50 violations per subject, split between 20 short violations and 30 long violations), and because the network violations often yield stimuli with large recency values (for example, 69\% of violations yield stimuli with recency greater than 10), we separate our data based on ranges of recency values that provide an approximately even distribution of violations (see Supplementary Fig. 4a). After separating the data by recency, we estimate the effects of short and long violations using a mixed effects model of the form `$\text{RT}\sim \log(\text{Trial}) + \text{Target} + \text{Top}\_\text{Dist} + (1 + \log(\text{Trial}) \,|\, \text{ID})$'. We note that, in comparison to the model used in Sec. \ref{violations}, we have removed the mixed effect of Top\_Dist because the filtered datasets are not large enough to provide a significant estimate. In Supplementary Fig. 4a, we see that, within each recency range other than recency greater than 40, at least one of either the short or long violations generates a significant increase in reaction times relative to standard transitions. Additionally, in Supplementary Fig. 4b, we filter the violations data by the number of appearances of the current stimulus in the previous 10 trials. Network violations yield significant increases in reaction times across all conditions other than short violations with zero appearances in the last 10 trials. Together, these results demonstrate that the effects of network violations cannot simply be explained by recency, therefore suggesting that subjects maintain an internal representation of the transition structure.

\begin{figure}[t!]
\centering
\includegraphics[width = .65\textwidth]{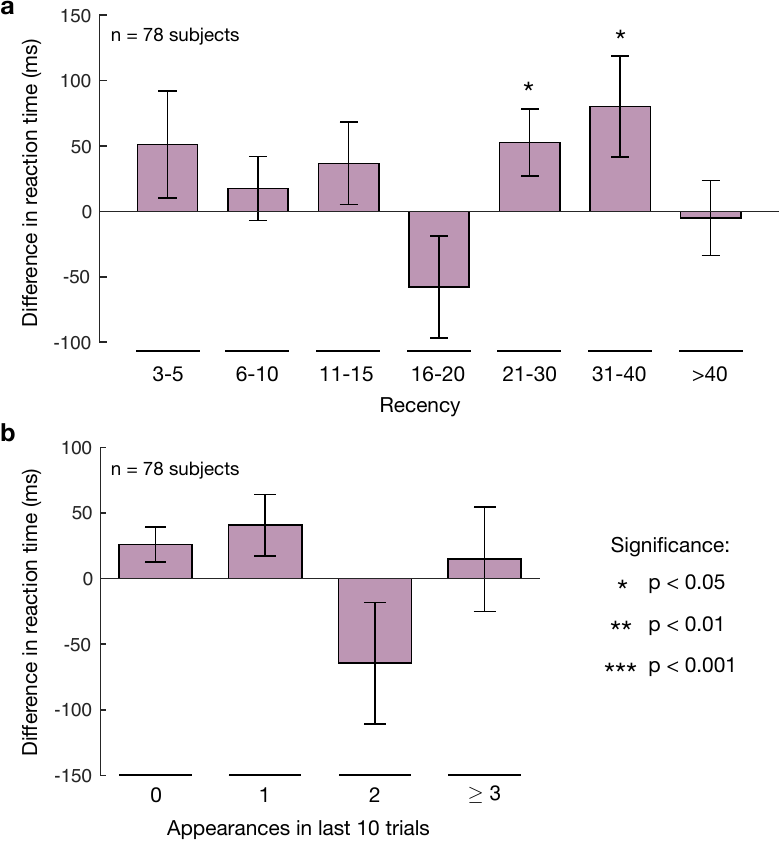} \\
\raggedright
\myfont\textbf{Supplementary Fig. 5: Comparing short versus long network violations while controlling for recency.}
\captionsetup{labelformat=empty}
{\spacing{1.25} \caption{\myfont \textbf{a}, Difference in reaction times between short and long network violations after controlling for the recency of stimuli. We find significant increases in reaction times for long violations in the recency ranges 21-30 and 31-40. \textbf{b}, Difference in reaction times between short and long network violations after controlling for the number of times the current stimulus has appeared in the previous 10 trials. Effect sizes (represented by bar plots), standard deviations (represented by error bars), and \textit{p}-values are estimated using mixed effects models. The results are measured for all 78 subjects that observed random walks with violations in the ring graph. Source data are provided as a Source Data file.}}
\end{figure}

We repeat the above analysis to measure the difference in reaction times between short and long violations while controlling for recency. We observe a significant increase in reaction times for long violations relative to short violations in two of the seven recency ranges (Supplementary Fig. 5a). However, we do not report a significant difference in reaction times after controlling for the number of appearances of stimuli in the previous 10 trials (Supplementary Fig. 5b). We remark that, given the noisy nature of reaction times and the small number of measurements per subject, the large standard deviations in our estimates are not surprising. Nevertheless, these results provide tentative evidence that recency alone cannot explain the difference in reaction times between long and short network violations.

\section{The forgetting of stimuli cannot explain network effects}
\label{forgetting}

In the derivation of our model, the central mathematical object is the memory distribution $P(\Delta t)$, which represents the probability that a person recalls the stimulus that occurred at time $t - \Delta t$ instead of the stimulus that they were trying to recall at time $t$. Generally, this memory distribution is intended to reflect the erroneous shuffling of past stimuli in a person's memory. Alternatively, one could imagine errors in memory that reflect the forgetting of past stimuli altogether, a process that has recently been shown to impact human reinforcement learning\cite{Collins-01, Collins-02} and to facilitate flexible and generalizable decision making.\cite{Richards-01} Here we argue that this second form of cognitive errors -- that is, the simple forgetting of stimuli -- cannot explain the higher-order network effects described in the main text.

Consider a sequence of stimuli reflecting a random walk of length $T$ on a network defined by the transition matrix $A$, where $A_{ij}$ represents the probability of transitioning from stimulus $i$ to stimulus $j$. Given a running tally $n_{ij}(T)$ of the number of times each transition has occurred, we recall that the most accurate prediction for the transition structure is given by the maximum likelihood estimate $\hat{A}^{\text{MLE}}_{ij}(T) = n_{ij}(T)/\sum_k n_{ik}(T)$. Now suppose that a human learner forgets each observed transition at some fixed rate. On average, this process of estimating $A$ after forgetting some number of transitions uniformly at random is equivalent to estimating $A$ at some prior time $T_{\text{eff}}$. In other words, forgetting observed transitions at random simply introduces additional white noise into the transitions estimates $\hat{A}^{\text{MLE}}_{ij}(T)$. As discussed in the main text, maximum likelihood estimation provides an unbiased estimate of the transition structure, and therefore cannot explain the fact that people's representations depend systematically on higher-order network organization. Similarly, the addition of white noise to $\hat{A}^{\text{MLE}}(T)$ will also yield an unbiased (although less accurate) estimate of the transition structure. Therefore, while the forgetting of past stimuli plays an important role in a number of cognitive processes,\cite{Richards-01, Collins-01, Collins-02} this mechanism cannot be used to explain the higher-order network effects observed in human experiments and predicted by our model.

\section{Gradient of RMS error with respect to inverse temperature $\beta$}
\label{gradient}

Given a sequence of nodes $x_t$, we recall that our prediction for the reaction time at time $t$ is given by $\hat{r}(t) = r_0 + r_1 a(t)$, where $a(t) = \hat{A}_{x_{t-1},x_t}(t-1)$ is the predicted anticipation of node $x_t$. The gradient of the RMS error $\text{RMSE} = \sqrt{\frac{1}{T}\sum_t (r(t) - \hat{r}(t))^2}$ with respect to the inverse temperature $\beta$ is given by
\begin{equation}
\label{dRMSE}
\frac{\partial \text{RMSE}}{\partial \beta} = \frac{-r_1}{T}\frac{1}{\text{RMSE}} \sum_t \left(r(t) - \hat{r}(t)\right) \frac{\partial a(t)}{\partial \beta},
\end{equation}
where the derivative of the anticipation is given by
\begin{equation}
\frac{\partial \hat{A}_{ij}(t)}{\partial \beta} = \frac{1}{\sum_k \tilde{n}_{ik}(t)}\Bigg(\frac{\partial \tilde{n}_{ij}(t)}{\partial \beta} - \hat{A}_{ij}(t)\sum_{\ell} \frac{\partial \tilde{n}_{i\ell}(t)}{\partial \beta}\Bigg).
\end{equation}
Recalling Eq. (8) from the main text, the derivative of the transition counts can be written
\begin{equation}
\frac{\partial \tilde{n}_{ij}(t)}{\partial \beta} = \sum_{t' = 1}^{t-1}\sum_{\Delta t = 0}^{t'-1} \frac{\partial P_{t'}(\Delta t)}{\partial \beta} \left[i = x_{t'-\Delta t}\right]\left[j = x_{t'+1}\right],
\end{equation}
where $P_{t'}(\Delta t)$ represents the probability of accidentally remembering the node $x_{t'-\Delta t}$ instead of the target node $x_{t'}$. Taking one more derivative, we have
\begin{equation}
\label{dP}
\frac{\partial P_{t'}(\Delta t)}{\partial \beta} = P_{t'}(\Delta t)\Bigg(-\Delta t + \sum_{\Delta t' = 0}^{t'-1} P_{t'}(\Delta t') \Delta t'\Bigg).
\end{equation}
Taken together, Eqs. (\ref{dRMSE})-(\ref{dP}) define the derivative of the RMS error with respect to the inverse temperature $\beta$, thus completing the description of our gradient descent algorithm.

\section{Connection to the successor representation}
\label{successor}

In the limit of an infinitely-long sequence of nodes, we showed in the main text that the transition estimates in our model take the following concise analytic form,
\begin{equation}
\hat{A} = (1-e^{-\beta})A(I-e^{-\beta}A)^{-1},
\end{equation}
where $A$ is the true transition structure, $\beta$ is the inverse temperature in our memory distribution, and $I$ is the identity matrix. Interestingly, this equation takes a similar form to the successor representation from reinforcement learning,
\begin{equation}
M = A(I - \gamma A)^{-1},
\end{equation}
where $\gamma$ is the future discount factor, which tunes the desired time-scale over which a person wishes to make predictions.\cite{Sutton-01, Gershman-01} Put simply, starting at some node $i$, the successor representation $M_{ij}$ counts the future discounted occupancy of node $j$. Identifying $\gamma = e^{-\beta}$, we notice that the successor representation is equivalent to an unnormalized version of our transition estimates. Moreover, the same mathematical form crops up in complex network theory, where it is known as the communicability between nodes in a graph.\cite{Estrada-01, Estrada-02, Garvert-01}

The relationship between the transition estimates in our model and the successor representation is fascinating, especially given the marked differences in the concepts that the two models are based upon. In our model, people attempting to learn the one-step transition structure $A$ instead arrive at the erroneous estimate $\hat{A}$ due to natural errors in perception and recall. By contrast, given a desired time-scale $\gamma$, the successor representation defines the optimal prediction of node occupancies into the future.\cite{Sutton-01, Gershman-01} Interestingly, the successor representation has been linked to grid cells and abstract representations in the hippocampus,\cite{Stachenfeld-01, Garvert-01} decision making in reward-based tasks,\cite{Momennejad-01, Russek-01} and the temporal difference and temporal context models of learning and memory.\cite{Sutton-01, Gershman-01, Howard-01} The successor representation assumes that humans are making predictions multiple steps into the future; however, our results show that a similar mathematical form can instead represent a person who simply attempts to predict one step into the future, but misses the mark due to natural errors in cognition. This biologically-plausible hypothesis of erroneous predictions highlights the importance of thinking carefully about the impact of mental errors on human learning.\cite{Richards-01, Collins-01, Collins-02}

\newpage

\section*{References}

\bibliographystyle{naturemag}
\bibliography{GraphLearningBib}

\end{document}